\documentclass[%
 reprint,
superscriptaddress,
 amsmath,amssymb,
 aps,
 prx
]{revtex4-2}

\usepackage[displaymath]{lineno}  
\usepackage{graphicx}
\usepackage{amssymb}
\usepackage{amsmath}
\usepackage[utf8]{inputenc}
\usepackage{cancel}
\usepackage[usenames,dvipsnames]{color}
\usepackage{psfrag}
\usepackage{epsfig}
\usepackage{bbm}
\usepackage{bm}
\usepackage{dsfont}
\usepackage{soul}
\usepackage{colonequals}
\usepackage{float}
\usepackage{enumitem}
\usepackage{appendix}
\usepackage{lmodern}
\usepackage[T1]{fontenc}
\usepackage{physics}                                                
\usepackage{hyperref}
\usepackage[capitalize]{cleveref}
\usepackage{xcolor}			
\usepackage{siunitx} 
\usepackage{flafter}

\DeclareSIUnit{\molar}{M}

\newcommand{\ccite}[1]{\IfSubStr{#1}{,}{Refs.~}{Ref.~}\cite{#1}} 

\newcommand{\tf}{t_\mathrm{f}}
\newcommand{\lf}{\lambda_\mathrm{f}}
\newcommand{\x}{x}
\newcommand{\dotx}{\dot{x}}
\newcommand{\ddotx}{\ddot{x}}
\newcommand{\xb}{x_b}
\newcommand{\dotxb}{\dot{x}_b}
\newcommand{\lam}{\lambda}

\definecolor{DarkRed}{RGB}{153,0,0}
\definecolor{DarkGreen}{RGB}{0,153,0}
\definecolor{DarkBlue}{RGB}{0,0,153}

\newcommand{\out}[1]{ }

\newcommand{\dt}{\mathrm{d}t}

\newcommand{\la}{\left\langle}
\newcommand{\ra}{\right\rangle}

\graphicspath{{Figures/}}
\begin{document}

\nolinenumbers

\preprint{APS/123-QED}

\title{Universal symmetry of 
optimal control
at the microscale}

\author{Sarah A. M. Loos}\email{sl2127@cam.ac.uk}
\address{Department of Applied Mathematics and Theoretical Physics, Centre for Mathematical Sciences, University of Cambridge, Cambridge, United Kingdom}
\author{Samuel Monter}
\address{Faculty of Physics, University of Konstanz, Konstanz, Germany}
\author{Felix Ginot}
\address{Faculty of Physics, University of Konstanz, Konstanz, Germany}
\author{Clemens Bechinger}\email{clemens.bechinger@uni-konstanz.de}
\address{Faculty of Physics, University of Konstanz, Konstanz, Germany}

\date{\today}

\begin{abstract}
Optimizing the energy efficiency of driving processes provides valuable insights into the underlying physics and is of crucial importance for numerous applications, from biological processes to the design of machines and robots. Knowledge of optimal driving protocols is particularly valuable at the microscale, where energy supply is often limited. Here we investigate experimentally and theoretically the paradigmatic optimization problem of moving a potential carrying a load through a fluid, in a finite time and over a given distance, in such a way that the required work is minimal. An important step towards more realistic systems is the consideration of memory effects in the surrounding fluid, which are ubiquitous in real-world applications. Therefore, our experiments were performed in viscous and viscoelastic media, which are typical environments for synthetic and biological processes on the microscale. Despite marked differences between the protocols in both fluids, we find that the optimal control protocol and the corresponding average particle trajectory always obey a time-reversal symmetry. We show that this symmetry, which surprisingly applies here to a class of processes far from thermal equilibrium, holds universally for various systems, including active, granular, and long-range correlated media in their linear regimes. The uncovered symmetry provides a rigorous and versatile criterion for optimal control that greatly facilitates the search for energy-efficient transport strategies in a wide range of systems. Using a machine learning algorithm, we demonstrate that the algorithmic exploitation of time-reversal symmetry can significantly enhance the performance of numerical optimization algorithms.
\end{abstract}

\maketitle

The increasing quest of energy-efficient machines and processes is not limited to macroscopic length scales. Due to ongoing miniaturization, micro- and nanoscopic machines are within reach and there is great need for optimal control strategies that enable their efficient operation \cite{wang2018recent,woods2001energy}. This is also motivated by biological engines (molecular motors) which achieve remarkable efficiencies even at high frequencies and in the presence of liquid environments \cite{grover2016transport,toyabe2011thermodynmic}. At first glance, this seems to be in contradiction with conventional heat engines  where maximal efficiency is only reached for infinitely slow, i.e., quasi-static motion \cite{blickle2012realization,martinez2016brownian,schmiedl2008efficiency,Esposito2010efficiency,holubec2018cycling}. Recent studies of molecular machines suggest that by regulating their power according to the external resistance, excessive dissipation can be avoided \cite{gupta2022optimal,sivak2012thermodynamic}. 

The energy-efficient operation of nanomachines is a specific example of a finite-time, (near-)optimal driving process~\cite{schmiedl2007optimal,sivak2012thermodynamic,van2023thermodynamic,proesmans2020finite,bechhoefer2021control,aurell2011optimal,aurell2012refined,gomez2008optimal,abreu2011extracting}. Such processes are not only important for small-scale robotic applications \cite{li2017micro, wang2018recent, woods2001energy} but also in various other fields like
plant physiology~\cite{king1982graded}, molecular biology \cite{schiebinger2019optimal}, as well as classical \cite{proesmans2020finite} and quantum information processing~\cite{caneva2009optimal,van2023thermodynamic}. As a generic example of a finite-time optimal process, theoretical studies have considered the most energy-efficient dragging of a micron-sized particle through a viscous environment. Surprisingly, this is achieved by a non-steady forcing with jump discontinuities at the beginning and end \cite{schmiedl2007optimal,gomez2008optimal}.

So far, experimental and theoretical studies on optimal driving processes have mainly considered viscous environments, which remain in equilibrium in the presence of a driven particle. Typically, however, molecular motors or microrobots operate within more complex surroundings. Common environments are viscoelastic media (e.g., intra-cellular plasma \cite{mao2022passive}, blood \cite{brust2013rheology}, 
polymeric gels \cite{schamel2014nanopropellers,lin2011polymer}, micellar solutions \cite{rehage1988rheological}, or colloidal suspensions \cite{vanderwerff1989linear,naegele1998linear}) which do not remain in equilibrium during operation due to their slowly relaxing microstructure. This leads to pronounced memory effects, i.e., non-Markovian behavior of driven particles in such media, which are absent in purely viscous (memory-free) liquids \cite{khan2029optical,cao2023memory}.

Here, we experimentally and theoretically investigate the optimal control problem of translating optical tweezers containing a Brownian particle over a given distance within a given time, with minimum average work done on the particle. In viscous liquids, our experiments confirm previously predicted driving discontinuities at the beginning and the end of the protocol and a constant driving power in between \cite{schmiedl2007optimal}. For viscoelastic fluids, however, no constant driving regime is found due to the presence of memory effects. Astonishingly, despite the fundamentally different optimal protocols in viscous and viscoelastic fluids, we find that both the mean particle trajectory and the protocol for optimal control exhibit a time-reversal symmetry. Starting from a generalized Langevin equation, we theoretically prove that this symmetry generally arises irrespective of the specific memory kernel or noise property, in both the overdamped and underdamped regime, as long as all forces are linear. Therefore, time-reversal symmetry allows a clear distinction to identify optimal control in various systems including 
granular \cite{sarracino2010irreversible} and active \cite{argun2016non,granek2022anomalous} media, or fluids with hydrodynamic backflow \cite{franosch2011resonances} or anomalous diffusion \cite{burov2008critical}. 
Beyond their implications for the optimal operation of micromachines in complex environments, our findings suggest that the use of symmetry considerations is a powerful tool to significantly expedite optimization problems.

\section{Experimental setup}\label{sec:Experimental Details}
\begin{figure}[H] 
	\begin{centering}            \includegraphics[keepaspectratio = true,width = 8cm]{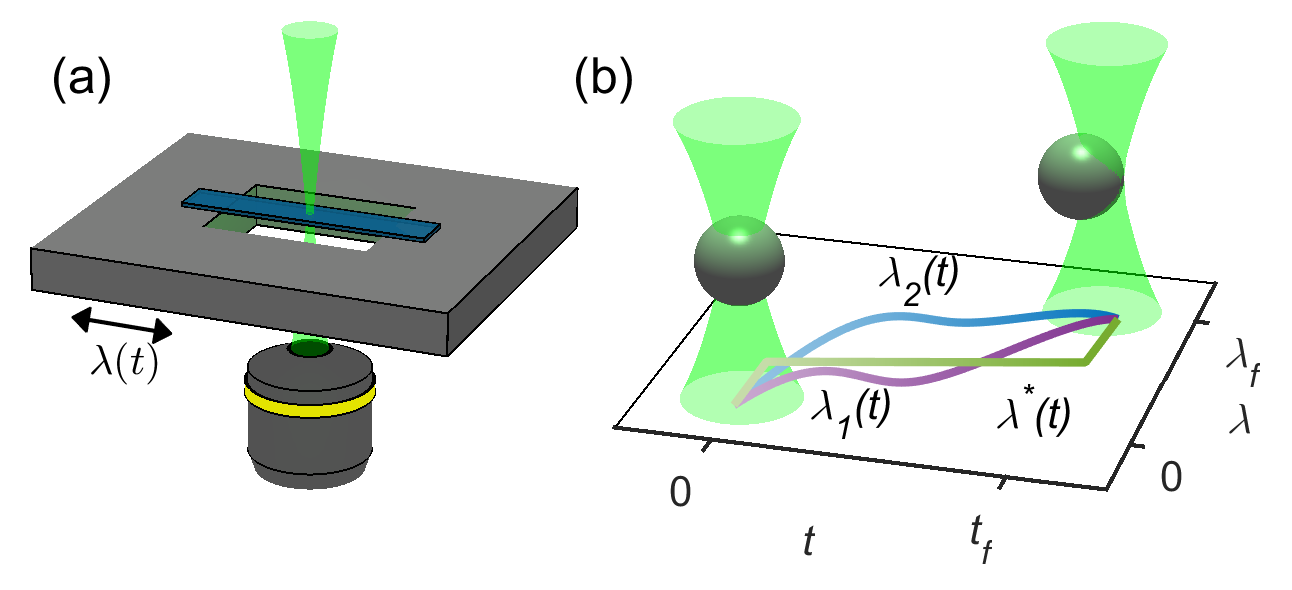}
		\caption{
  (a) A colloidal particle is trapped in optical tweezers. Relative movement of the trap center $\lambda$ is achieved by translating a piezo-actuated sample stage. (b) The protocol $\lambda(t)$ describes the transfer of the trap center from its initial $\lambda(0) = 0$ to its final position $\lambda(\tf) = \lf$ during the time $\tf$. $\lambda^*$ corresponds to the protocol of minimal work.}
        \label{fig:sketch}
	\end{centering}
\end{figure}
Our experiments were performed using spherical silica particles (diameter $\approx 2.7\,\unit{\micro\meter}$) suspended in a fluid contained in a sample cell with 100~\unit{\micro\meter} height.
In addition to a purely viscous water-glycerol mixture (1:1), we used a viscoelastic solution composed of cetylpyridinium chloride monohydrate (CPyCl) and sodium salicylate (NaSal) (for details see  Appendix \ref{sec:expmeth}). At the used concentration of 8\,\unit{\milli\molar} a worm-like micellar micro-network is formed \cite{cates1990statics} whose relaxation time was experimentally determined to $\tau_b \approx 17\,\unit{\second}$ (see Appendix \ref{sec:measproc}). A laser beam (532\,nm) is focused with an objective (NA = 1.45, 100x) into the mid-plane of the sample cell (see Fig.~\ref{fig:sketch}a) where it creates a parabolic optical potential
{\par\nobreak\noindent} 
\begin{align}
V(X,\lambda) = \frac{\kappa}{2} (X-\lambda)^2 .
\end{align} 
Here, $\kappa$ denotes the trap stiffness, $X$ the particle position and $\lambda$ the time-dependent trap center location, which is the control parameter. Experimentally, a dynamical variation of $\lambda$ with time $t$ was realized by translating the sample cell relative to a static optical trap. The positions of the particle and the trap were determined by digital video microscopy with a spatial and temporal resolution of 5\,\unit{\nano\meter} and 10\,ms, respectively.
All experiments were performed at a sample temperature of 25~\unit{\celsius}. We applied protocols that shift the trap between an initial $\lambda(t=0)=0$ and a final position $\lambda(\tf)= \lf$ within the time interval $\tf$ (see Fig.~\ref{fig:sketch}b). After each protocol we waited for the system to fully equilibrate. Thus, $\langle X(0) \rangle=\lambda(0)=0$ for each run.

The work $W$ exerted on the particle during such protocols is determined according to \cite{jarzynski1997nonequilibrium,crooks1998nonequilibrium} 
{\par\nobreak\noindent} 
\begin{align} \label{def:W}
  W[\lambda,X] &= \int_{0}^{\tf} \dot{\lambda}  \frac{\partial V}{\partial \lambda} \mathrm{d}t = \kappa \int_{0}^{\tf}  \dot{\lambda}(\lambda - X) \mathrm{d}t\, .
\end{align}
Because $W$ fluctuates between individual protocols due to thermal noise, each measurement was repeated at least 100 times to yield mean values $\la W \ra$ with a relative statistical error $<1\,\%$. 
In the following, we determine the optimal protocol $\lambda^*$ which requires the smallest $\la W \ra$.

\section{Optimal control in viscous liquids}
\begin{figure*}[ht] 
	\begin{centering}
		\includegraphics[keepaspectratio = true]{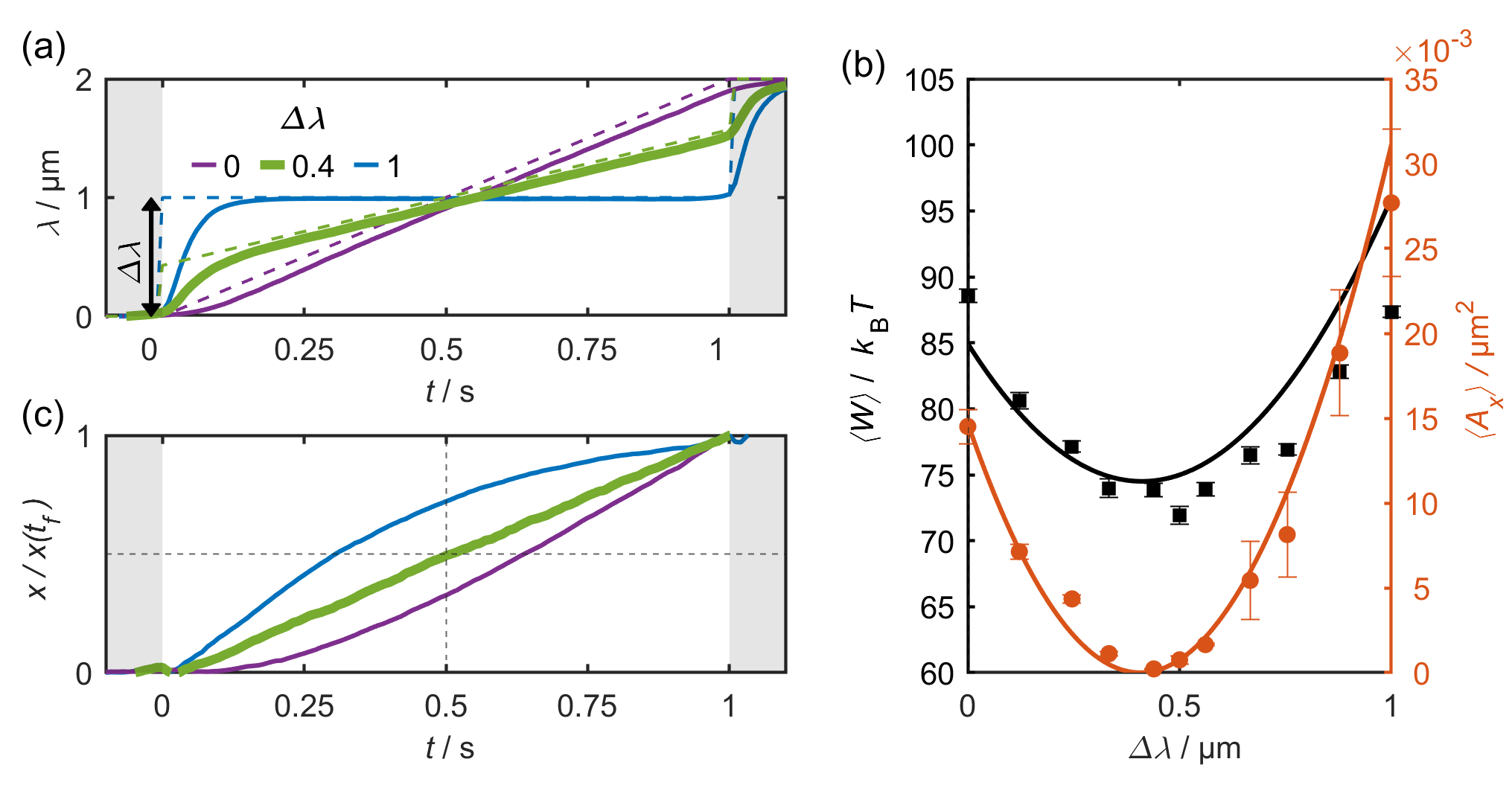}
        \caption{ 
        (a) Experimental (solid lines) protocols for the motion of an optical trap which is displaced by $\lf = 2$\,\unit{\micro\meter} within the time $\tf$=1\,\unit{\second}. Different protocols are characterized by their jump height $\Delta\lambda$. The dashed lines correspond to protocols with ideal infinitely fast jumps at start and end of the protocol. The optimal protocol (green) $\lambda^*$ corresponds to $\Delta\lambda^* = 0.4$\,\unit{\micro\meter} (see Eq.~\eqref{lambdaoptMarkov} in the Appendix~\ref{sec:modeloptvis}). 
        (b) Experimentally measured mean work $\la W\ra$ (black) and asymmetry parameter $\langle A_x \rangle$ (orange) as function of $\Delta\lambda$ (symbols). The minima of $\la W\ra$ and $\la A_x\ra $ are positioned at the optimal protocol $\lambda^*$ in agreement with theoretical predictions (solid lines). The fact that $\la A_x \ra\approx 0$ for $\lambda^*$ suggests that both $\lambda$ and $x$ are time symmetric for optimal control. Error bars correspond to the s.e.m.. 
        (c) Average particle trajectories $x=\la X \ra$ normalized by the position at the end of the protocol $x(\tf)$ corresponding to the protocols shown in (a).}
        \label{fig:panelV}
	\end{centering}
\end{figure*}

For viscous liquids, the optimal protocol $\lambda^*$ is known to exhibit symmetric jumps at $t=0$ and $t=\tf$ and a constant trap velocity in between~\cite{schmiedl2007optimal} (see Appendix \ref{sec:modeloptvis}). Such protocols are fully quantified by the jump height $\Delta \lambda$.  Figure~\ref{fig:panelV}a shows experimental (solid lines) and ideal (dashed lines) protocols for $\Delta \lambda = 0, 0.4$, and $1\,\unit{\micro\meter}$ with $\lf = 2\,\unit{\micro\meter}$ and $\tf = 1\,\unit{\second}$. Since the acceleration of the translational stage is finite, instantaneous jumps cannot be perfectly realized in experiments. Compared to ideal protocols, this leads to deviations at the beginning and the end of $\lambda(t)$ and slightly increases the time during which the optical trap exerts work on the particle. To take this effect into account, when calculating the work [Eq.~\eqref{def:W}] the upper integration limit was increased accordingly (see Appendix \ref{sec:measproc}).

Figure~\ref{fig:panelV}b shows the experimentally determined mean work $\la W \ra$ (black symbols), which exhibits a minimum at $\Delta \lambda^* \approx 0.5\,\unit{\micro\meter}$. Despite the above mentioned experimental limitations in realizing ideal jumps, our data are in good agreement with theoretical predictions 
for protocols with
infinitely fast jumps. To match the theoretical prediction with the experiments, we used the relaxation time $\tau_0 = 0.35\,\unit{\second}$ calculated from the measured equilibrium distribution and the mean square displacement of the particle in the static trap (see Appendix \ref{sec:measproc}). Without any adjustable parameters, the theory then gives $\Delta \lambda^* = 0.41\,\unit{\micro\meter}$ (black line)~\cite{schmiedl2007optimal}, which agrees well with the experimental findings. 

Mean particle trajectories $x = \la X \ra$ corresponding to the executed protocols are plotted in Fig.~\ref{fig:panelV}c. Interestingly, we find that only near the optimal protocol, 
$x$ obeys time-reversal symmetry, which can be formally written as, 
{\par\nobreak\noindent} 
\begin{align}\label{eq:Symmetry}
	f(t)= - f(\tf-t) + f(\tf)\,
\end{align}
with $f\equiv x$.
This becomes even clearer
when the deviation from time-reversal symmetry of $x$ is quantified. For this purpose we introduce the asymmetry parameter 
{\par\nobreak\noindent} 
\begin{equation}
A_f = \frac{1}{\tf}\sum_{t=0}^{\tf/2} \left[f(t) + f(\tf-t) - 2\, f\left({\tf}/{2}\right)\right]^2  \Delta t\, \label{eq:Ax}
\end{equation}
with $\Delta t$ corresponding to the temporal resolution. $A_f$ is zero for time-symmetric functions $f$ and increases with increasing asymmetry [note that $f(0) = 0$, namely $\lambda(0)=0$ by choice and without loss of generality, and $x(0)=\lambda(0)=0$ due to the relaxed initial condition]. For an illustration of $A_f$ we refer to Fig.~\ref{fig:visA} in the Appendix.

Figure~\ref{fig:panelV}b shows the asymmetry parameter $A_{x}$ defined in \eqref{eq:Ax} evaluated for $f=x$, versus $\Delta\lambda$; which exhibits a minimum at the experimentally obtained ideal protocol at $\Delta \lambda^* \approx 0.5\,\unit{\micro\meter}$ (symbols). 
In agreement with the experimental findings, the theoretically predicted trajectories for the perfectly symmetric protocols (with instantaneous jumps) also lead to coinciding minima of $\la A_x \ra$ and $\langle W\rangle$.

\section{Symmetry proof} 

As will be demonstrated below, the occurrence of time-reversal symmetry of both $x^*$ and $\lambda^*$ is a distinctive and universal property of a wide class of optimal processes.
To prove this surprising observation, we start from a general generalized Langevin equation (GLE) \cite{kubo1966fluctuation,mori1965transport,zwanzig2001nonequilibrium,waigh2005microrheology,snook2006langevin,doerries2021correlation}
{\par\nobreak\noindent} 
\begin{align}\label{GLE}
m\ddot{X}+	\int_{-\infty}^{t}\Gamma(t-t')\dot{X}(t') \mathrm{d}t' =  
 -\nabla V
 + \nu(t)\, ,
\end{align}
with an arbitrary memory kernel $\Gamma$  
and zero-mean noise $\nu$, and particle mass $m$. We further assume that the potential $V$ is parabolic (as being the case for an optical trap), rendering the GLE to be linear in $X$ and $\lam$. Notably, no further restrictions are applied to 
$\Gamma$ or $\nu$. The latter may be a colored or a non-Gaussian noise, and we do not restrict to systems obeying the fluctuation-dissipation relation $\Gamma(|t-t'|)\propto \langle \nu(t)\nu(t')\rangle$. Within such assumptions, Eq.~\eqref{GLE} describes the dynamics of a broad class of systems, including granular media \cite{sarracino2010irreversible}, glasses in their ergodic regime
\cite{cummins1997dynamics,elizondo2019glass}
, linear viscoelastic fluids~\cite{waigh2005microrheology}, polymer condensates \cite{jawerth2020protein}, 
particles in bacterial baths \cite{argun2016non},
and even tracers in molecular fluids  \cite{straube2020rapid}. 
The above GLE also applies to memory-free, viscous liquids by using delta-correlated kernels. Furthermore, the overdamped limit, which is the relevant regime in our experiments, is readily included by taking the limit $m\to 0$.

Based on Eq.~\eqref{GLE}, we obtain general implications and non-implications between optimality and time-reversal symmetry of the corresponding protocol $\lambda$ and the mean observable (trajectory) $x =\la X \ra$. The results are summarized in Fig.~\ref{fig:diagram}. Here, we outline the main idea of the proof, which is described in detail in Appendix~\ref{sec:theoproofs}.

First, all non-implications can be proved by simple counterexamples (see Appendix \ref{sec:counterex}). For instance, the non-implication
\textit{time-reversal symmetry of $\lambda$} $\nRightarrow$ \textit{time-reversal symmetry of $x$} becomes immediately obvious by considering an entirely linear protocol $\lambda(t)=(\lf/\tf) t$ 
which leads in a purely viscous, overdamped system to an exponentially relaxing trajectory which is clearly not symmetric under time reversal. 

Next, by rewriting the work as a functional of either $\x$ or ${\lam}$, only, one can show that optimality directly implies time-reversal symmetry.  
Because the confining potential is quadratic, both in $\lambda$ and $x$, the functionals are also quadratic, from which the symmetry directly follows. 
Concretely, as we explicitly show in the Appendix~\ref{App:opt-implies-symmetry-x} based on Eqs.~\eqref{def:W} and \eqref{GLE}, the mean work  can be expressed as the functional of $x$
{\par\nobreak\noindent} 
\begin{align}\label{G-functional}
	\la W \ra
	=&
 \int_{0}^{\tf}\!\! \!   \int_{0}^{t}\! \Gamma(t-t') \dot{x} (t) \dot x(t')  \mathrm{d}t'  \dt  + m[\dot{x}(\tf)^2]/2
\nonumber \\&
  +
 \kappa[x(\tf)- \lf]^2/2 + \mathcal{C} 
 ,
\end{align}
where $\mathcal{C}$ encapsulates terms that are independent of the process during $t\in[0,\tf]$ and therefore irrelevant for the optimization or symmetry property. 
To derive Eq.~\eqref{G-functional} we have further made use of the fact that $\la W \ra$ is independent of the noise level, which is a consequence of the linearity of the GLE.
By applying appropriate coordinate transformations,
one can show that the functional in Eq.~\eqref{G-functional} is invariant under time reversal of $x$, irrespective of the particular memory kernel $\Gamma$. Specifically, 
any trajectory $x$ and its time-reversed image $\widetilde x(t) \equiv - x(\tf-t)+ x(\tf)$ yields the same average work, $\langle W[x] \rangle \equiv \langle	W[\widetilde{x}]  \rangle $. 
Given the quadratic form of the work functional, we expect only a unique extreme, such that $x^* \equiv \widetilde x^*$, implying that
optimal trajectories are always time-reversal symmetric. Similarly, by explicitly making use of causality of the stochastic process, we can express the mean work as a quadratic functional of $\dot{\lam}$
convoluted with the response function $\Phi$ [see Eq.~\eqref{condition_optimalProtocols} in Appendix~\ref{App:opt-implies-symmetry-x}]. 
Analog arguments then also imply that optimal protocols $\lam^*$ must be time-reversal symmetric.

Finally, to show the reversed implication, we reexamine the work functional~\eqref{G-functional}. Optimal trajectories $x^*$ are characterized by a vanishing variation of the work $\delta \!\la W \ra$ with respect to all variations of $x^*$ that satisfy $\delta x(0)\!=\!\delta x(\tf)\equiv 0$. Using variational calculus and some integral manipulations, we find that $\delta \!\la W \ra = 0$, if and only if
{\par\nobreak\noindent} 
\begin{align}\label{condition_optimalTrajectories}
	 \frac{\mathrm{d}}{\mathrm{d}t} \left[\int_{0}^{\tf} \Gamma(|t-t'|) \dot{x}^* (t') \mathrm{d}t'  \right] = 0.~~~\forall t\in [0,\tf]\,.
\end{align}
See Appendix~\ref{App:Symmetry-x-implies-opt} for a derivation.
In addition, for all time-symmetric protocols, one can show that 
any time-symmetric mean solution $x(t)$ of Eq.~\eqref{GLE} for $t \in [0,\tf]$ fulfills the equality
{\par\nobreak\noindent} 
\begin{align}\label{eq:intermediate-step-proof-3}
	\int_0^{\tf}\Gamma(|t-t'|) \dot{x}(t')\mathrm{d}t'
	& = \kappa[\lf - x(\tf)] \, 
\end{align}
as given in Eq.~\eqref{eq:intermediate-step-proof-3-M} in Appendix~\ref{App:Symmetry-x-implies-opt}. Realizing that the equality~\eqref{eq:intermediate-step-proof-3} readily implies the condition \eqref{condition_optimalTrajectories}, it follows that $\la W \ra$ is optimal whenever $x$ and $\lam$ are both time-reversal symmetric.

Overall, we have thus shown that the simultaneous time-reversal symmetry of both $x$ and $\lam$ is a necessary and sufficient condition for optimal control. Importantly, this holds for \textit{all} linear GLE models.
 
\begin{figure}
	\centering
	\includegraphics[width=0.5\textwidth]{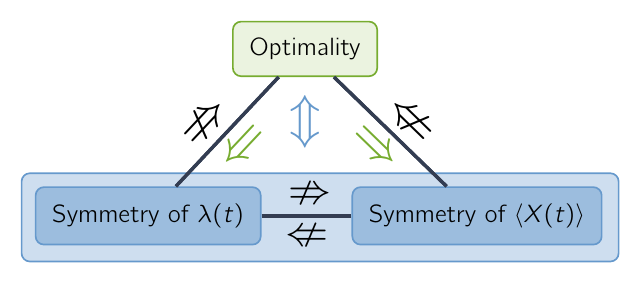}
	\caption{
     Diagram showing implications ($\Rightarrow$) and non-implications ($\nRightarrow$) between time-reversal symmetry [defined in Eq.~\eqref{eq:Symmetry}] of the protocol $\lambda$ and mean particle trajectory $\la X\ra=x$, and optimality with respect to the work required, for control processes in linear media. Optimal control is characterized by time-reversal symmetry of both $\lambda$ and $\la X\ra$.}
	\label{fig:diagram}
\end{figure}

\section{Optimal control in viscoelastic fluids} 

To experimentally test the generality of the predicted
symmetry and to explore the generic features of optimal control in non-Markovian environments, we 
performed dragging experiments in a viscoelastic micellar fluid. To realize experimental variations around $\lam^*$, we first obtain a theoretical prediction regarding the optimal protocol. 
Our calculations are based on previous studies, demonstrating that the memory kernel of micellar viscoelastic solutions is well-described by a single exponential decaying with the bath stress-relaxation time $\tau_b$~\cite{rehage1988rheological,waigh2005microrheology,jawerth2020protein}. We further treat the dynamics in the overdamped limit, because the inertial effects in our experimental system are negligible. As a consequence, the motion of 
a colloidal particle within such a Maxwell fluid is experimentally observed to be in agreement with two coupled overdamped equations \cite{caspers2023how,ginot2022recoil}
{\par\nobreak\noindent} 
\begin{align}
	\tau_p \,\dot{X} &= -\frac{\kappa}{\kappa_b}(X-\lambda) - (X-X_b)  +  \xi_p  \, ,\label{LEx} \\ %
	\tau_b \,\dot{X}_b &= -(X_b-X)  + \xi_b \, .\label{LEy}
\end{align}
Here, $X$ and $X_b$ correspond to the positions of the colloid and a fictitious bath particle connected by a harmonic spring with stiffness $\kappa_b$, and $\xi_p$, $\xi_b$ are uncorrelated Gaussian white noises with variances ${{2k_\mathrm{B}T}}\tau_i/{\kappa_b}$, $i=p,b$, respectively, and $\tau_p:=\gamma/\kappa_b$.

Expressing the work as functional of $\la X \ra$ and $\la X_b\ra$ using Eqs.~\eqref{def:W} and \eqref{LEx}, and incorporating Eq.~\eqref{LEy} as dynamical constraint via a Lagrange multiplier, we construct an appropriate cost functional \cite{bechhoefer2021control}, for which the Euler-Lagrange equations yield $x^*, \lam^*$ (see Appendix \ref{sec:protVE}).

\begin{figure*}[ht] 
	\begin{centering}
		\includegraphics[keepaspectratio = true]{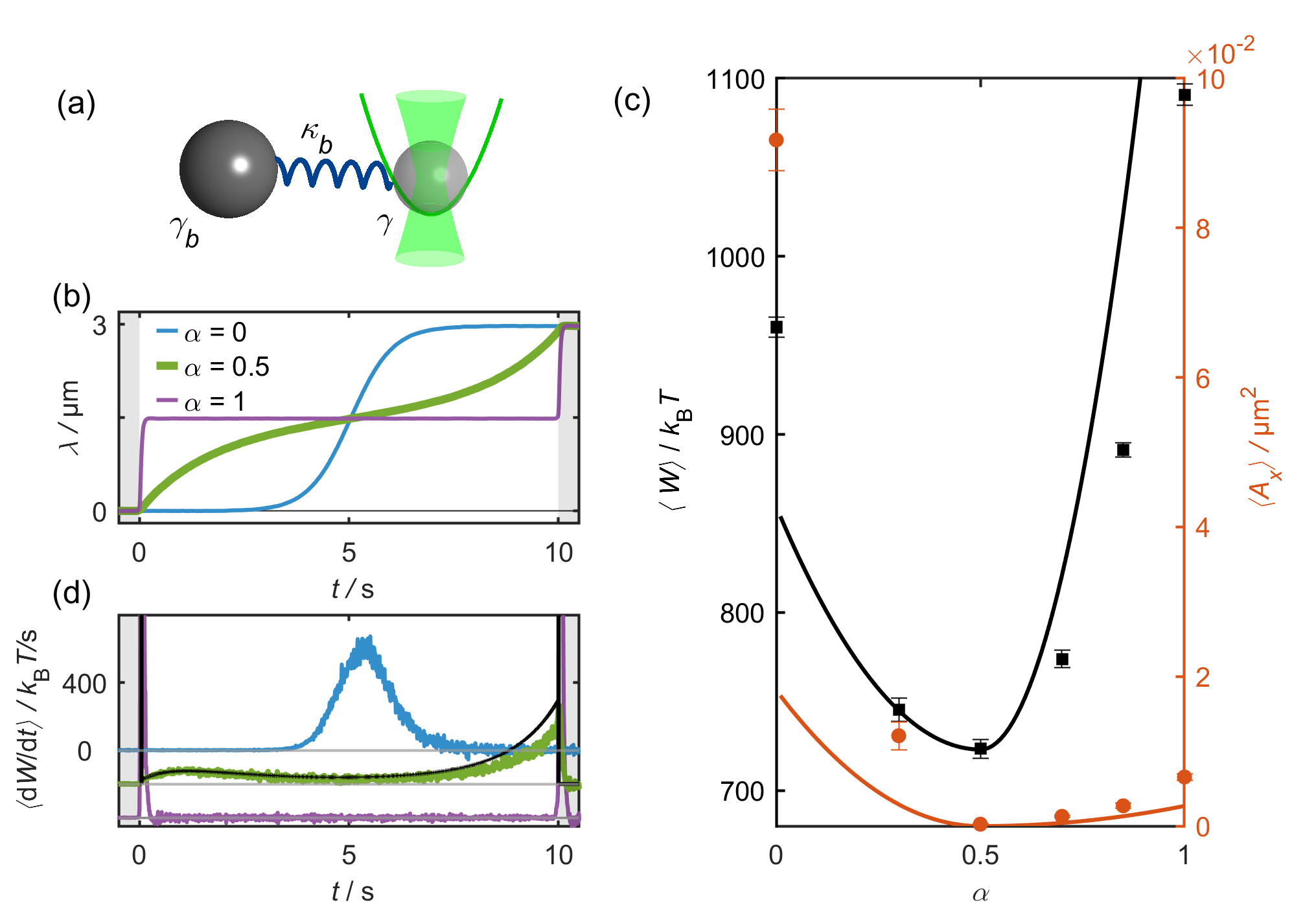}
		\caption{
        (a) The sketch visualizes the used Maxwell model with fictitious bath particle with friction $\gamma_b$, coupled to the tracer particle with friction $\gamma$ with a linear spring with stiffness $\kappa$. Only the tracer feels forces from the optical potential.
        (b) Different time-reversal symmetric control protocols $\lambda(t)$ were executed, which are a linear combination of a $\tanh$-function (blue line), the predicted optimal protocol (green line) and a step-like function (purple line) weighed by a parameter $\alpha$ (for details see Appendix \ref{sec:measproc}). We depict the corresponding particle trajectories in Fig.~\ref{fig:protNtrajVEopt} in the Appendix.
        (c) The experimentally measured mean work $\la W\ra$ (black symbols) shows a distinct minimum close to $\alpha = 0.5$ that is for the predicted optimal protocol.
        Theoretical predictions (black line) using the Maxwell model [Eqs.~\eqref{LEx},\,\eqref{LEy}], which are offset by $-290\, k_\mathrm{B}T$, are in good qualitative agreement with the experimental data. The asymmetric contribution $A_x$ (orange) quantifies the deviation of the mean trajectory $x$ from time-reversal symmetry and exhibits a minimum at the same position as $\la W\ra$. Error bars again show the s.e.m.. (d) Measured time resolved power $\langle \mathrm{d}W/\mathrm{d}t \rangle$ traces for the protocols shown in (b). Individual protocols are offset to negative values by $200\, k_\mathrm{B}T$ for better readability, gray lines indicate the corresponding y-axis origin. Protocols including a jump show strong peaks at beginning and end. The work trace for the optimal protocol (green line) shows a complex non-linear and non-monotonic shape. 
        The matching curve from numerically solving the equation of motion (black line) is in good agreement with the experiment.}
        \label{fig:panelVE}
	\end{centering}
\end{figure*}

The theoretically predicted optimal protocol, as realized in our experiment in a viscoelastic micellar solution, is shown as green solid line in Fig.~\ref{fig:panelVE}b. 
As expected, $\lam^*$ and $\x^*$ display time-reversal symmetry.
Similar to the Markovian case (even though hardly visible here), 
the optimal protocol exhibits jumps at the beginning and end. However, in non-Markovian systems $\lambda^*$ contains no linear parts. Only in the quasi-static limit $\tf \rightarrow \infty$ where the memory kernel has decayed to zero, the protocol converges to the Markovian case and becomes linear.

To experimentally test the theoretically calculated $\lambda^*$, we have varied the protocol around this prediction. Variations around $\lambda^*$ were achieved by a variation parameter $\alpha$, with $\alpha=0.5$ corresponding to $\lambda^*$ (see Appendix \ref{sec:measproc}). Exemplarily, Fig.~\ref{fig:panelVE} shows experimental protocols for three different values of $\alpha$, all of them exhibiting time-reversal symmetry as suggested by our theoretical considerations.
Because the friction in the viscoelastic fluid is larger than that in the viscous water-glycerol mixture, the mean optical trap velocity had to be reduced
by setting $\lf = 3\;\unit{\micro\meter}$ and $\tf = 10\;\unit{\second}$. Figure~\ref{fig:panelVE}c shows the measured averaged work $\la W \ra$ versus $\alpha$, indicating a pronounced minimum at $\alpha=0.5$. 
In addition, the asymmetry
of the mean trajectories $\la A_x \ra$ exhibits a minimum for the optimal protocol, which is in excellent agreement with the prediction of time-reversal symmetry of the optimal process independent of the solvent's memory effects.

The appearance of time-reversal symmetry of both $\lambda$ and $x$ for the optimal solution in non-Markovian systems 
(see Appendix \ref{sec:measproc}, Fig. \ref{fig:protNtrajVEopt})
is particularly astonishing in view of the strongly time-asymmetric nature of the overall process, which starts in thermal equilibrium and ends in a state where the particle within the optical trap and the surrounding are anisotropic and out of equilibrium. This becomes visible in the time-resolved power $\la \mathrm{d}W/\mathrm{dt} \ra$, shown in Fig.~\ref{fig:panelVE}c for different time-symmetric protocols, which is clearly not time-symmetric even for $\lambda^*$ (green curve). In fact, $\la \mathrm{d}W /\mathrm{d}t\ra$ is strongly time-asymmetric and non-monotonic; and the largest contribution to the applied work occurs towards the end of the optimal protocol. This modulated power input differs significantly from the Markovian case, where optimal control is characterized by constant power between the jumps 
(see Appendix \ref{sec:dataeval}, Fig.~\ref{fig:worktraceV}).
 
The nonlinear behavior of $\lam^*$ and asymmetry of $\la \mathrm{d}W \ra$ in a viscoelastic fluid can be rationalized by energetic considerations. At the beginning of the protocol, the fluid is isotropic and fully relaxed. In this regime, the particle can be initially dragged with relatively low energetic costs even at high trap velocities. With increasing particle displacement, however, the viscoelastic microstructure becomes increasingly distorted, thereby accumulating elastic energy and thus increasing the particle resistance against the trap motion. Therefore, to avoid an excessive increase of work, the trap velocity should be rather slow. However, since the protocol must be completed within $\tf$, the trap velocity cannot be slow over the entire protocol. This conflict is resolved by slowing down the trap for $t<\tf/2$ and accelerating it afterward. With this strategy, elastic energy stored in the fluid becomes largest towards the end of the protocol, 
when the trap has already stopped (at $\lf$) and therefore exerts no more work on the particle.
We remark that these arguments are in good agreement with the observation that molecular motors modulate their driving in accordance to their resistance \cite{sivak2012thermodynamic}.

\section{Application in Machine learning}
\begin{figure}[ht] 
	\begin{centering}
      \includegraphics[keepaspectratio = true]        {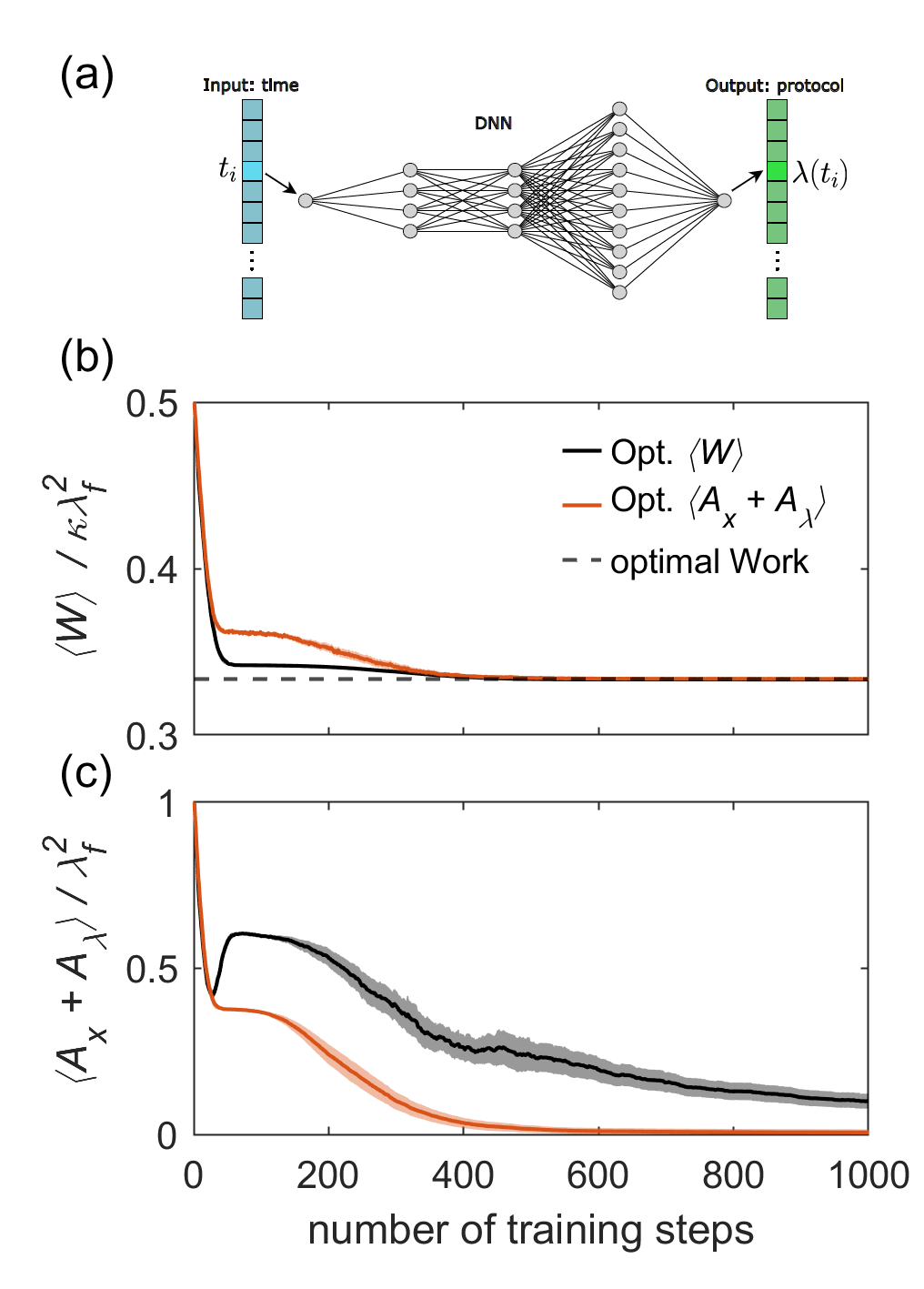}
		\caption{
          We train a deep neural network (DNN) with the objective of minimizing the work $\la W\ra$ (black lines), or the overall asymmetry $A_x+A_\lam$ (orange lines). (b) Shows the work, (c) shows the total asymmetry during training. The lines depict the averages of 100 independent training runs, the shaded regions indicate standard derivations. Both objectives lead to successful learning. Training on $  A_x+A_\lam  $ has the crucial advantage that the globally optimal value $ A_x+A_\lam  \to 0$ is \textit{a priori} known. As illustrated in (a) the network has 5 layers of \{1,4,4,10,1\} nodes and training is done by a Monte Carlo algorithm, similar to 
          Ref.~\cite{whitelam2023demon}, see Appendix \ref{sec:ML} for details. 
          }
        \label{fig:ML}
	\end{centering}
\end{figure}

Finally, we aim to address an important practical implication of the uncovered symmetry in the context of computational optimization. In this field, recent progress has been made by incorporating machine learning techniques \cite{engel2023optimal,whitelam2023demon,whitelam2023train}.  
To demonstrate the benefit of the symmetry property in this context, we trained a deep neural network to find the optimal control protocol in a viscous overdamped system, {following an algorithm similar to~\cite{whitelam2023demon}}. 
Instead of training the network with the objective of minimizing the work (Fig.~\ref{fig:ML}, black line), we also trained with the objective of minimizing the asymmetries of protocol and mean trajectory $A_x+A_\lam$ (Fig.~\ref{fig:ML}, orange line). 
In both cases, the learned protocol converges to the optimal one, demonstrating that the asymmetry parameter can be used as an alternative (or additional) cost functional for the optimization. 
Moreover, for control in more complex environments, for which generally no analytical solution is available, the asymmetry parameter offers the essential advantage that its value at the global optimum ($A_x+A_\lam \equiv 0$) is always and \textit{a priori} known. Therefore, unlike the work, the asymmetry reveals whether the true work optimum has been reached.

\section{Conclusion \& Outlook}
With experiments and theoretical calculations, we have studied the optimal control of an optical trap dragging a colloidal particle through viscous (memory-free) and viscoelastic (non-Markovian) baths. Despite the very different response of colloidal particles in both systems, we find that the optimal protocols and the corresponding mean particle trajectories always exhibit time-reversal symmetry. Remarkably, such symmetry which is a hallmark of thermal equilibrium, reappears in these systems far away from equilibrium, but only when driven in the most energy-efficient way. The symmetry is rather universal, 
applying to all systems 
in the regime where the
dynamical equations are approximately linear in the system's variable and protocol parameter. It is thus expected to be valid also in systems with long-ranged memory,  
such as 
active fluids \cite{argun2016non,granek2022anomalous} or fluids with hydrodynamic backflow \cite{franosch2011resonances}. 
This symmetry property of optimal control is of immediate relevance for the efficient operation of nanomachines, low energy swimming mechanisms of micro-robotic systems 
but also 
energy-efficient information processing~\cite{dago2023adiabatic,proesmans2020finite}. 
In machine learning,   
numerical and exact optimization,
the {a priori} knowledge of the uncovered time-reversal symmetry provides a strong constraint for the applicable function space and will, thus, significantly improve optimization algorithms.

\begin{acknowledgments}
SL acknowledges funding through the postdoctoral fellowship of the Marie Skłodowska-Curie Actions (Grant Ref. EP/X031926/1) undertaken by the
UKRI, and through the Walter Benjamin Stipendium (Project No. 498288081) from the 
Deutsche Forschungsgemeinschaft (DFG). CB acknowledges funding through the DFG, Grant No. SFB 1432 - Project ID 4252172. We thank Mike Cates, Robert Jack, and \'Edgar Rold\'an for valuable discussions.
\end{acknowledgments}

\subsection*{Author contributions}
CB, SM and FG designed the experiment. The experimental data were taken and analyzed by SM and discussed together with FG and CB who supervised the project. SL performed the analytical calculations, derived the theoretical proof and implemented the machine learning model.
SL, SM, and CB wrote the manuscript.


\appendix

\section{Experimental Methods}
\label{sec:expmeth}
\subsection{Optical trap setup}
The optical tweezers setup used consists of a 532\,nm laser (Coherent Verdi V2) that is amplitude modulated with an acousto-optic deflector (AOD, AA opto-electronics DTSXY-400). The laser power in front of the AOD was adjusted to 250\,mW using a $\lambda/2$-plate and a polarizing beam splitter. As a result, the laser power after the AOD remained below 100\,mW. A 100x objective (Olympus Apochromat MPLAPON-Oil 100x NA=1.45) was used to focus the laser beam into the sample cell. To manipulate the sample position, a 3-axis piezo-driven stage (piezoconcept LT3) was used. 
Its position was controlled using the analog input with a signal supplied with a sampling rate of 5\,kHz by a PCIe card (National Instruments PCIe-6351). At the same time, the actual position of the stage was measured from the analog output signal. Differences between the set and actual positions occur when the stage position is adjusted fast. The sample temperature was controlled by resistive heating of the sample stage and microscope objective (Okolab). The temperature was kept at 25$^\circ$C throughout all experiments. For video microscopy, the same microscope objective was used to give an image on a digital camera (Basler ace 2 a2A3840-45umPRO). Videos were acquired during the experiment at a frame rate of 100 frames per second. 

\subsection{Sample preparation}

The viscoelastic solution was prepared by dissolving equimolar amounts of cetylpyridinium chloride monohydrate (CPyCl) and sodium salicylate (NaSal) in milipor water. The solution was stirred overnight to ensure equilibration of the micellar network. In this study, we used an 8~\unit{\milli\molar} solution.\\
Sample solutions were prepared by dispersing colloids (2.73~\unit{\micro\meter} $\mathrm{SiO}_2$, microParticles GmbH) in either a water-glycerol mixture or the viscoelastic solution using an ultrasonic bath.
Samples were prepared by filling glass capillaries with an inner diameter of 100~\unit{\micro\meter} (CM Scientific) with the sample solution. The capillaries were closed with a combination of wax and epoxy resin. 
After filling the capillary, the samples were equilibrated in the measurement setup until the measured mean work shows no drift anymore.\\

\section{Measurement Procedures}
\label{sec:measproc}
All protocols consist of moving the trap center $\lambda$ according to a specific protocol over a distance $\lf$ in a time $\tf$. 
The trap stiffness $\kappa$ remains constant. 
Translation of the trap center was technically realized by translating the sample with a piezo-driven stage relative to the static trap. 
Even though this makes the temporal resolution of the optical trap position relative to the sample slower compared to e.g. an acoustic-optical deflector (AOD), our approach has several advantages: first, the optical path of the trap is not altered during translation which avoids spatial changes to its shape and stiffness. In addition, a resting optical trap in the reference frame of the camera makes it possible to considerably reduce the field of view of the camera and thus to achieve a high temporal resolution of the particle's motion.
In a single measurement run, the protocol was executed in both the forward and backward directions consecutively, ultimately ending up in the same trap position.
Before and after each individual protocol, the system was allowed to relax for a specified time interval $t_{rel}$. 

\subsection{Variation of optimal protocols in experiments}
\label{sec:MethVarProt}
To experimentally investigate the optimality of protocols predicted by the theory, the protocols were systematically varied as a function of one parameter.

\subsubsection{Protocol variations for viscous samples}

\begin{figure}[h]
    \includegraphics[scale = 1]{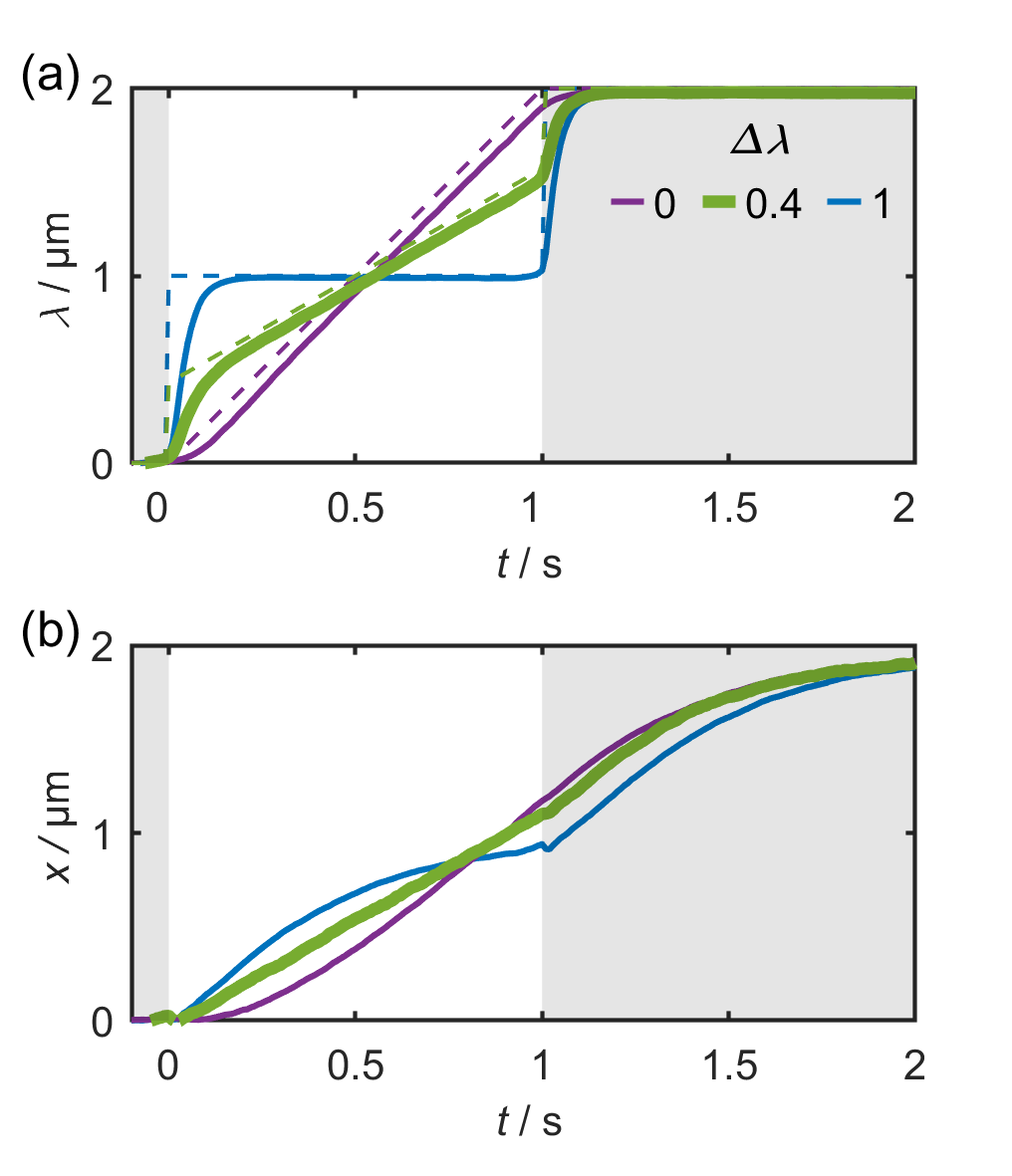}
    \caption{
    (a) Protocols and (b) and mean particle trajectories from experimental data for the viscous system, for the same set of experiments as shown in Fig.~\ref{fig:panelV}. 
    Different from Fig.~\ref{fig:panelV}, we here also show the relaxation period of the mean particle trajectory after $\tf$.
    $\Delta \lambda=0.4$ yields the optimal control. Only for this case, both $\lambda$ and $x$ exhibit time reversal symmetry on $0<t<\tf$. The protocol duration and total trap displacement were $\tf = 1\,\unit{\second}$ and $\lf = 2\,\unit{\micro\meter}$. }
    \label{fig:relaxationV}
\end{figure}

For the (Markovian) case of a particle in a viscous fluid, the optimal protocol for the shifting of a trap center in a finite time is given by Eq.~\eqref{lambdaoptMarkov}.
The optimal protocol, which has symmetric jumps of height $\Delta \lambda^*$ at the beginning and end, can be seen as a member of the family of functions described by
{\par\nobreak\noindent} 
\begin{align}
    \lambda(0<t<\tf) &= \frac{\lf-2\Delta \lambda}{\tf}\,t + \Delta \lambda\,. \label{eq:lambdagen}
\end{align}
To check the optimality of $\lambda^*$, protocols defined by Eq.~\eqref{eq:lambdagen} with varying jump height $\Delta \lambda$ were executed. Examples comparing the theoretical given shape of the protocol and the realization in experiments together with resulting mean particle trajectories are shown in Fig.\ref{fig:relaxationV}. In this figure the relaxation of the particle after the protocol is finished can be observed.

\begin{figure}[b] 
    \includegraphics[keepaspectratio = true, scale = 1]{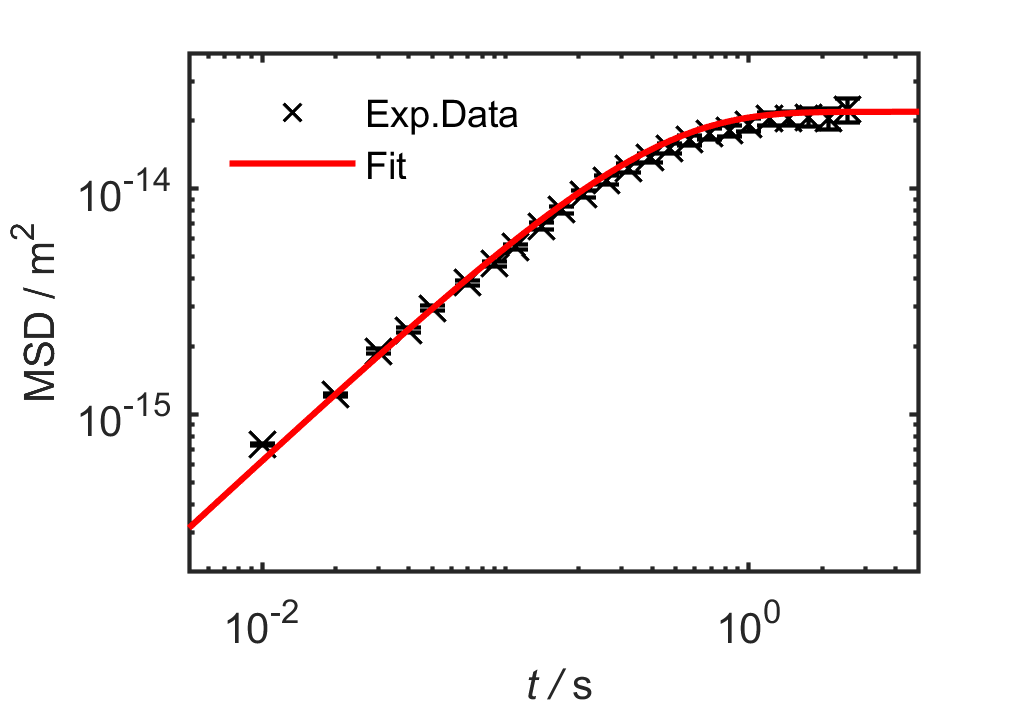}
		\caption{
        Mean-squared displacement (MSD) of the particle in the stationary trap to experimentally determine the friction coefficient $\gamma_0$. Symbols show the ensemble averaged data from experiments, errorbars correspond to the SEM. In full lines, the analytical expression for the MSD of a 2.73\,\unit{\micro\meter} colloid in a harmonic trap is plotted using the fitted parameters. The trap stiffness $\kappa$ is fitted from the positional distribution in equilibrium and $\gamma_0$ is derived afterward from the initial slop of the MSD \cite{giesler2021optical}.}
        \label{fig:MSDV}
\end{figure}

To calculate a prediction for the optimal jump height $\Delta\lambda^*$, it is necessary to know the relaxation time of the particle in the trap $\tau_0$.
The latter is given as the quotient of particle friction $\gamma_0$ and trap stiffness $\kappa$, both of which can be determined from equilibrium measurements~\cite{giesler2021optical}.
The trap stiffness $\kappa$ was extracted by fitting a parabolic function to the potential derived from the equilibrium distribution using the Boltzmann factor.
Using the derived trap stiffness $\kappa = 0.38\,\unit{\micro\newton\per\meter}$, the friction coefficient $\gamma_0$ was subsequently calculated from the initial slope of the mean square displacement 
(see Fig.~\ref{fig:MSDV}). This gave a result of $0.13\,\unit{\micro\newton\second\per\meter}$ for $\gamma_0$ and therefore $\tau_0 = 0.35\;\unit{\second}$.

\subsubsection{Protocol variations for viscoelastic samples}

The optimal protocol for the viscoelastic solution, i.e., the non-Markovian case, is more complex. In Appendix \ref{sec:protVE}, we describe how we theoretically obtain the exact and general expression for it. The optimal protocol depends on the system parameters $\gamma$, $\gamma_b$, $\kappa_b$ and $\kappa$, 
defined in Eqs.~\eqref{LEx}, \eqref{LEy}, which first needed be determined experimentally before the optimal protocol can be executed in the experiment.
 To this end, we first conducted test experiments. These consisted of shifting the trap with constant velocity $\lf / \tf$.
The average work $\la W_\mathrm{exp} \ra$ calculated from experimental data was compared to simulation data. The difference in average work between simulation and experiment was minimized using a gradient descent algorithm by adjusting $\gamma$, $\gamma_b$ and $\kappa_b$ in the simulation model. The resulting parameter set was used to calculate $\lambda^*$.

\begin{figure}[b]
    \includegraphics[scale = 1]{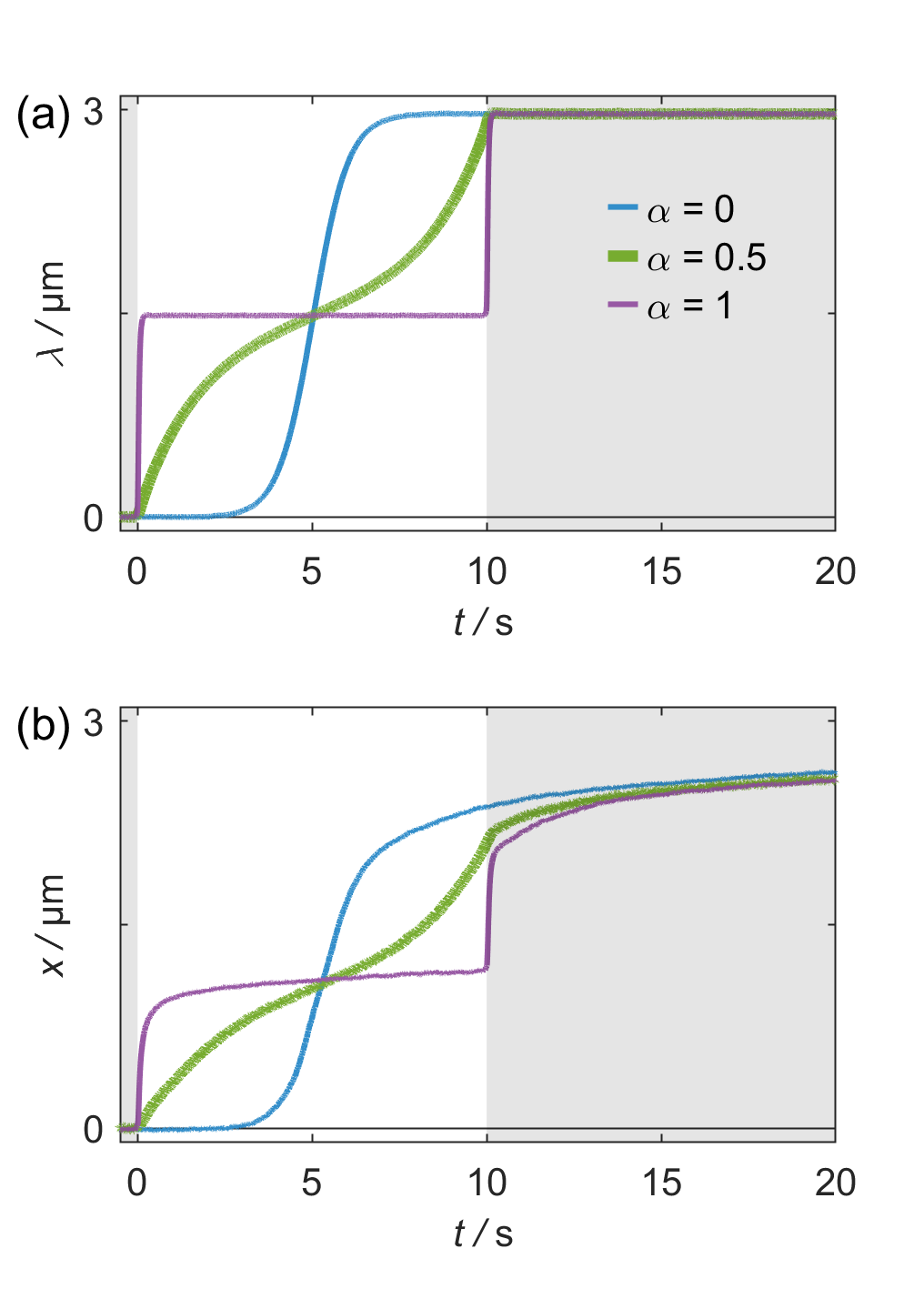}
    \caption{
    (a) Protocols and (b) mean particle trajectories from experimental data for the viscoelastic system, for the same set of experiments as shown in Fig.~\ref{fig:panelVE}. For the optical protocol ($\alpha = 0.5$), both $\lambda$ and $x$ exhibit time reversal symmetry on $0<t<\tf$. Protocol duration and total trap displacement were $\tf = 10\,\unit{\second}$ and $\lf = 3\,\unit{\micro\meter}$.}
    \label{fig:protNtrajVEopt}
\end{figure}

To enable a protocol variation in dependence of a single parameter (like in the Markovian case), we constructed a superposition of three different protocols in dependence of one parameter $\alpha$. 
For the three base functions we chose a $\tanh$-function [$\lam_T$, given in Eq.~\eqref{eq:lambdaNMtanh}], the optimal protocol [$\lam^*$, see Appendix \ref{sec:protVE}], and a step function featuring two symmetric jumps at the beginning and the end [$\lam_S$, given in Eq.~\eqref{eq:lambdaNMstep}].
These are linearly combined [see Eq.~\eqref{eq:lambdaNMvary}], i.e., added up, each one multiplied with an individual weighting factor $w_i \leq 1$, with i=T,S,O, ``O'' for ``optimal''.
The weighting factors can be expressed via single mixing parameter $\alpha$ according to Eqs.~\eqref{eq:w_O}-\eqref{eq:w_S}:
\begin{align}
    \lambda(t) &= w_\mathrm{T} \,\lambda_\mathrm{T}(t)  + w_\mathrm{S} \,\lambda_\mathrm{S}(t) + w_\mathrm{O} \,\lambda^*(t) \, ,\label{eq:lambdaNMvary}\\   
    \lambda_\mathrm{T}(t) &= \frac{\lf}{2}\Bigl[\tanh\Bigl(t-\frac{\tf}{2}\Bigr)+1\Bigr] \, , \label{eq:lambdaNMtanh}\\
    \lambda_\mathrm{S}(t) &= \begin{cases}
        0 &\text{for}\; \tau < 0\\
        \lf/2 &\text{for}\; 0 \leq \tau < \tf\\
        \lf &\text{for}\; \tau \geq \tf
    \end{cases} \, ,\label{eq:lambdaNMstep}\\    
    w_\mathrm{O} &= 1 - |2\alpha-1| \, ,\label{eq:w_O}\\
    w_\mathrm{T} &= \max(1-2\alpha,0)\, ,\label{eq:w_T}\\
    w_\mathrm{S} &= \max(2\alpha-1,0)\,. \label{eq:w_S}
\end{align}
By varying the mixing parameter $\alpha$, the individual contributions can be controlled. Notably $\alpha=0$ represents a mere $\tanh$ function [$\lambda(t) =\lambda_\mathrm{T}(t)$], $\alpha=0.5$ gives exactly the optimal protocol [$\lambda(t) =\lambda^*(t)$], and $\alpha=1$ the pure step function [$\lambda(t) =\lambda_\mathrm{S}(t)$]. Experimental realization of these edge cases with corresponding mean particle trajectories are shown in Fig.~\ref{fig:protNtrajVEopt}. In between these edge cases, the protocols are a superposition of the two respective functions. How we theoretically determine $\lam^*$ for the here considered viscoelastic case is explained Appendix \ref{sec:protVE}.

\subsection{Evaluation of experimental data}
\label{sec:dataeval}
The particle position is tracked by analyzing video recordings captured at a frame rate of 100\,fps using a custom tracking algorithm based on Ref.~\cite{trackingcode}. 
The position of the piezo-stage and the exposure active signal of the camera are recorded at 5 kHz using a PCIe card. 
This allows to synchronize the stage and particle positions {a posteriori}.

The work done during an individual protocol is calculated in the framework of stochastic thermodynamics using Stratonovich  integrals [see Eq.~\eqref{def:W}].
Since the trajectory is only sampled at finite intervals $\Delta t = 0.01\,\unit{s}$, the integral transforms into a sum:
{\par\nobreak\noindent} 
\begin{equation}
     W = \sum_{i = 1}^{N-1} \frac{\frac{\partial V}{\partial \lambda}\big|_{t_{i+1}} + \frac{\partial V}{\partial \lambda}\big|_{t_{i}} }{2} \Bigl[\lambda(t_{i+1}) - \lambda(t_{i}) \Bigr]\,.
\end{equation}
The calculated work was then averaged over multiple trajectories.

\begin{figure}[h]
    \includegraphics[keepaspectratio = true, scale = 1]{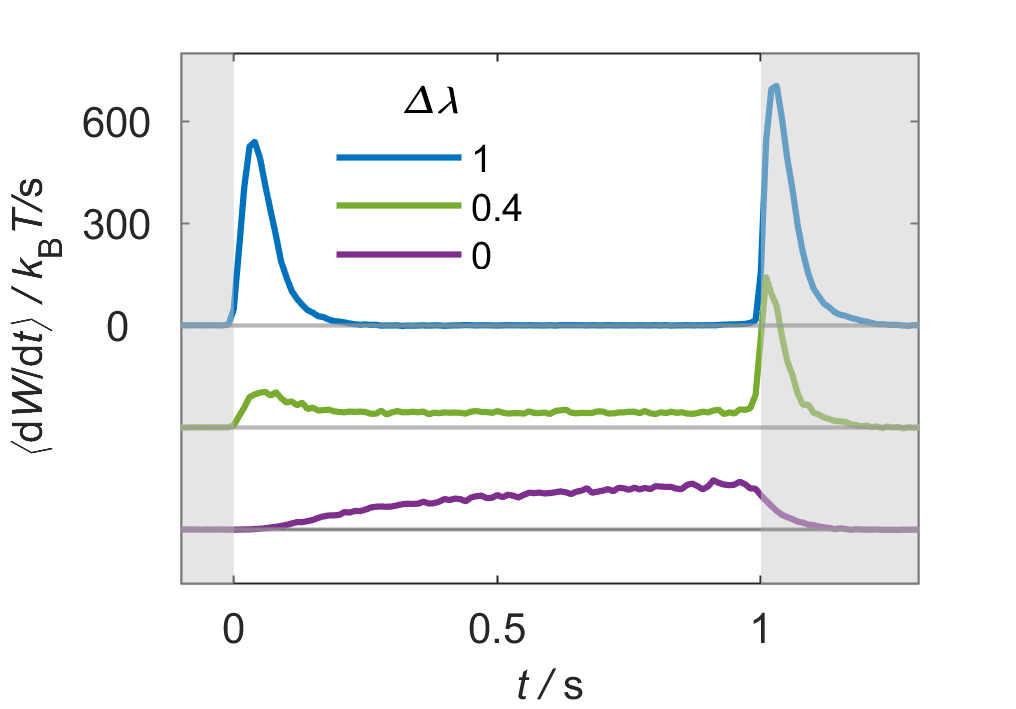}
    \caption{
    Measured work increment $\langle \mathrm{d}W/\mathrm{d}t \rangle$ traces over time for selected protocols executed in a viscous fluid. Individual protocols are offset to negative values by $300\,k_\mathrm{B}T$ for better readability, gray lines indicate the corresponding y-axis origin. The visible peaks result from jumps performed by $\lambda$. The work trace for the optimal protocol ($\Delta \lambda$ = 0.4\,\unit{\micro\meter}, green line in the middle) shows a constant work increment during the protocol execution.
    This is in agreement with theoretical prediction, which for the optimal protocol is $\la \mathrm{d}W /\mathrm{d}t\ra = (\Delta \lambda^*/\tau_0)^2 \tau_0,$ 
    for all $~0<t<\tf$.}
    \label{fig:worktraceV}
\end{figure}

Due to the inertia of the stage, it did not reach its final position $\lf$ at $\tf=1,\,10\;\unit{\second}$ but only about 0.1\,\unit{\second} later. To account for this contribution to the work, the time frame of work calculation was extended until $\dot{\lambda}\approx 0$. Similar considerations must be done for the asymmetry parameter $A_x$: The increased protocol time shifts the inflection point of the symmetry operation to slightly later times. To account for this, the time frame for calculating $A_x$ was extended to $\tf'$ so that the observed inflection point lies at $\tf'/2$. For this adjusted integration window, the asymmetry parameter of the protocol $A_\lambda$ is minimal.

\begin{figure}
    \centering
    \includegraphics{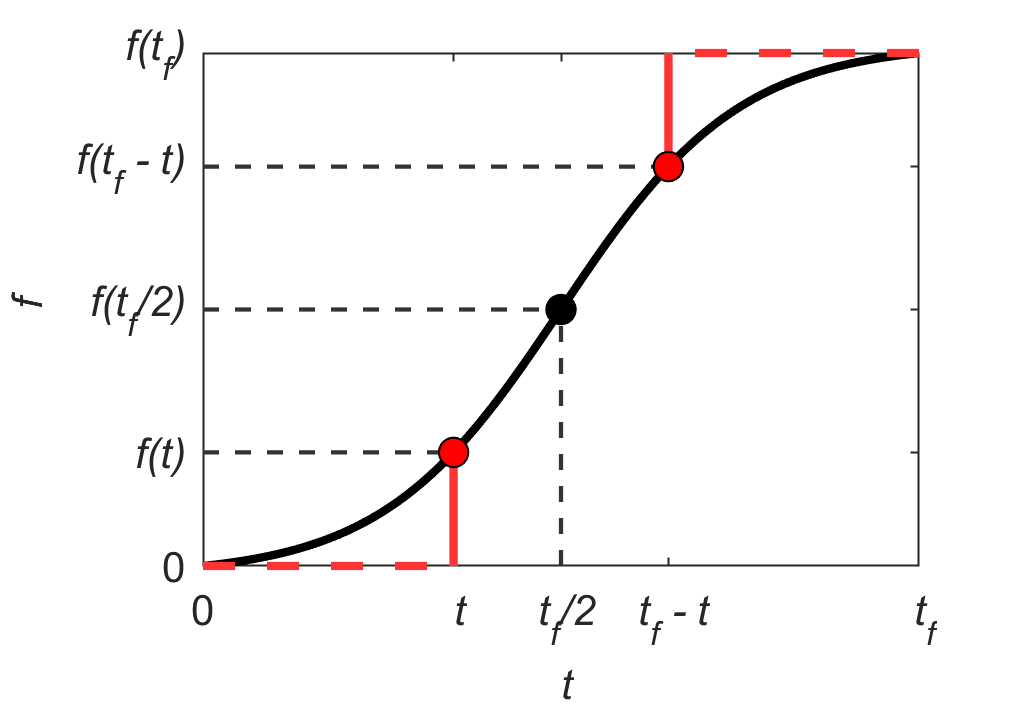}
    \caption{Visualization of the symmetry property and the asymmetry parameter $A_f$. Consider the difference between the two distances marked as red vertical lines [i.e., $f(t)$, and $f(t_f)-f(t_f-t)$]. For time-reversal symmetrical curves that are point-symmetrical around their center (black point), this operation results in zero, and $f(t_f)=2f(t_f/2)$. Any deviation from this symmetry leads to a remainder after subtraction of $f(t)$ and $2f(t_f/2)-f(t_f-t)$, which can be positive or negative. The absolute value increases with greater deviations. In order to always obtain a positive value, the result of the subtraction is therefore squared, giving rise to the definition of $A_f$ as given in Eq.~\eqref{eq:Ax}.}
    \label{fig:visA}
\end{figure}

\section{Model and theoretical prediction of optimal protocol in viscous fluid} 
\label{sec:modeloptvis}
The dynamics of the particle in the harmonic trap of stiffness $\kappa$ in the viscous fluid is described very accurately by the Markovian overdamped Langevin equation
{\par\nobreak\noindent} 
\begin{align}\label{eq:LE-Markov}
    \tau_0 \dot{X} = -(X-\lambda) + \xi 
\end{align}
with zero-mean, Gaussian white noise $\xi$ with
$\la \xi(t)\xi(t')\ra =2k_\mathrm{B}T\tau_0/\kappa\,\delta (t-t')$, where $\tau_0=\gamma_0/\kappa$ denotes the relaxation time in the trap, and $\gamma_0$ is the friction constant.
The seminal paper~\cite{schmiedl2007optimal} showed that the optimal protocol in this case has at the beginning ($t=0$) and end ($t=\tf$) two symmetric jumps of size $\Delta \lam^*={\lf}/({2+\tf/\tau_0})$, 
and in between increases linearly in time
{\par\nobreak\noindent} 
\begin{align}\label{lambdaoptMarkov}
\lambda^{*}(t) = \begin{cases}
0, &t=0\\
(\Delta \lam^*) (1+t/\tau_0), &0<t<\tf\\
\lf, &t=\tf
\end{cases}.
\end{align}
The corresponding optimal trajectories are fully linear: 
$\x^*(t) = \Delta \lam^*  (t/ \tau_0)$, and the optimal mean work takes the value $\la W^* \ra = (\Delta \lam^*) \gamma_0\lf/\tau_0.$ 
Thus, in the viscous fluid, optimal control is achieved, when particle and trap both exhibit a motion of the same constant speed $\Delta \lam^* / \tau_0$ for $0<t<\tf$, akin a steady state. In order to initiate this joint constant-speed motion, the trap needs to abruptly jump at time $t=0$, bringing the particle to the displacement from the trap center, where the relaxation force is constant and balanced with the trap's speed.

\section{Theoretical Proofs}
\label{sec:theoproofs}
\subsection{Counterexamples to show non-implications between symmetries and optimality}
\label{sec:counterex}
Here we briefly show all non-implications sketched in Fig.~\ref{fig:diagram}, by giving two counterexamples. To this end, we consider the case of an overdamped particle in a viscous fluid. First, while a fully linear protocol $\lambda(t)=(\lf/\tf) t$ possesses time-reversal symmetry, the corresponding mean particle trajectory violates it. 
This is seen in Fig. \ref{fig:panelV} (b), case $\Delta \lambda=0$ (purple curve), consistent with the theoretical prediction obtained by solving the LE~\eqref{eq:LE-Markov} for the linear protocol, which yields an exponential relaxation in the comoving reference frame: $x(t)=(\lf/\tf)[t-\tau_0(1-e^{-t/\tau_0})]$. We recall that $\tau_0=\gamma_0/\kappa$ denotes the trap relaxation time, and $\gamma_0$ the friction coefficient.
Thus \textit{symmetry of $\lam$} $\nRightarrow$ \textit{symmetry of $x$}. 
This counterexample also demonstrates that \textit{symmetry of $\lam$} $\nRightarrow$ \textit{minimal $\la W \ra$}.
As a second counterexample, consider the protocol $\lam(0)=0$, $\lam(t>0)=m (1 +t/\tau_0)$ with $m =\lf/( 1+\tf/\tau_0 )$, which has a single jump at $t=0$ and no jump at time $\tf$, and thus violates time-reversal symmetry. However, the corresponding mean trajectory is given by $x(t)=m' t$, and is thus time-symmetric; showing that \textit{symmetry of $x$} $\nRightarrow$ \textit{symmetry of $\lam$}, and that 
\textit{symmetry of $x$} $\nRightarrow$ \textit{minimal $\la W \ra$}.

\subsection{Proof of: Optimality $\Rightarrow$ Symmetry of $x$}\label{App:opt-implies-symmetry-x}
Here, we provide more details of the proof that optimality implies time-reversal symmetry of $x$.
As a first step, we express the work as a functional of $x$ only. To this end, we
first simplify Eq.~\eqref{def:W} by explicit integration of $\dot \lam \lam$ 
 and partial integration of $\dot \lam X$, taking into account the boundary conditions, which yields
{\par\nobreak\noindent} 
\begin{align} 
  W[\lambda,X] &= \kappa \left[\frac{\lf^2}{2} - X(\tf)\lf\right] +
  \kappa \int_{0}^{\tf} \lambda(t)\dot{X}(t)   \mathrm{d}t\, .
\end{align}
Next, substituting $\lam$ by
{\par\nobreak\noindent} 
\begin{align}
	\lam(t) = 
 \frac{m}{\kappa} \ddot{X}(t)+
\kappa^{-1} \int_{-\infty}^{t}\Gamma(t-t')\dot{X}(t') \mathrm{d}t' + X(t) - \kappa^{-1} \nu(t) \, ,
\end{align}
which follows from solving the GLE \eqref{GLE} for $\lam$, and taking the noise average, 
we find the functional
{\par\nobreak\noindent} 
\begin{align}\label{functionalW-x-with-noise}
	\la W \ra
	=&
 (\kappa/2) \la [X(\tf)-\lf]^2\ra -  (\kappa/2) \la X(0)^2\ra
 \nonumber \\&
 +(m/2) \la \dot{X}(\tf)^2 \ra - (m/2) \la\dot{X}(0)^2\ra
  \nonumber \\& 
 +\int_{0}^{\tf}\!\int_{-\infty}^{t}\!    \Gamma(t-t') \la \dot{X} (t) \dot X(t') \ra \mathrm{d}t'  \dt  
 \nonumber \\& 
 - 
    \int_{0}^{\tf}\! \la\dot X(t) \nu(t) \ra \mathrm{d}t
 \,.
\end{align}
We have used $\int_{0}^{\tf} \ddot{X}\dot{X}\dt=[\dot{X}^2]_0^{\tf}/2$.
Note that the first two terms of \eqref{functionalW-x-with-noise} is the increase of potential energy from time $0$ to $\tf$, and the third and fourth term is the increase of kinetic energy. To proceed further, we make use of the fact that the mean work is independent of the noise level. While this is directly not obvious from \eqref{functionalW-x-with-noise}, one can make it apparent as follows. Reconsidering Eq.~\eqref{def:W} and directly performing a noise average over it, we obtain $ \la W \ra = \kappa \int_{0}^{\tf}\dot{\lam}(\lam- x)\mathrm{d}t $, and conclude that $ \la W \ra$ depends only on the mean trajectory $x$, not on the fluctuations of $X$. Then, due to the linearity of the GLE, the temporal evolution of $x$ is fully independent of the noise, so that the same holds for $ \la W \ra $. To formally see the noise-independence of $x$, we also take the noise average of the GLE~\eqref{GLE}, which together with the initial condition yields 
\begin{align}\label{eq:noise-averagedGLE}
m\ddot{x}(t)+	\int_{0}^{t}\Gamma(t-t')\dot{x}(t') \mathrm{d}t' =  -\kappa [x(t)-\lambda(t)]\, ,
\end{align}
i.e., $x$ follows a deterministic (integro-differential) equation. Taken together, $\la W \ra$ itself is independent of the noise level. Hence, we are allowed to evaluate Eq.~\eqref{functionalW-x-with-noise} in the noise-free limit. For vanishing noise level, statistical dependencies between dynamical quantities disappear, so that $\langle \dot{X}(t)\dot{X}(t')\rangle \to \langle \dot{X}(t)\rangle \langle\dot{X}(t')\rangle =\dot{x}(t)\dot{x}(t')$ and $\langle \dot{X}(t)\nu(t)\rangle \to \la \dot{X}(t)\ra \la\nu(t)\ra \to 0$, simplifying \eqref{functionalW-x-with-noise}. Finally, we encapsulate in the constant $\mathcal{C}$ all terms of  Eq.~\eqref{functionalW-x-with-noise} that are independent of the process during $t\in[0,\tf]$ (and thus play no role in the optimization), allowing us to tighten the bounds of integration, and leading to 
{\par\nobreak\noindent} 
\begin{align}\label{G-functional-M}
	\la W \ra [x] 
	=&
 \int_{0}^{\tf}\!\!   \int_{0}^{t}\! \Gamma(t-t') \dot{x} (t) \dot x(t')  \mathrm{d}t'  \dt   +m[\dot{x}^2(\tf)]/2
 \nonumber \\&
 +
 \kappa[x(\tf)- \lf]^2 /2 + \mathcal{C} ,
\end{align}
as given in the main text in Eq.~\eqref{G-functional}. 

This functional is invariant under time reversal. 
To show this, we first consider only the integral term of $\la W \ra [x]$ and perform the coordinate transformations $t' = \tf-t''$, and $t = \tf-\tilde t$, which leads to
{\par\nobreak\noindent} 
\begin{align}
	&\int_{0}^{\tf}    \int_{0}^{t} \Gamma(t-t') \dot{x} (t') \dot x(t)  \mathrm{d}t'  \dt
		\nonumber \\
	&=
		\int_{0}^{\tf}    \int_{\tilde{t}}^{\tf} \Gamma(t''-\tilde{t}) \dot{x} (\tf-t'') \dot x(\tf -\tilde t)  \mathrm{d}t'' \mathrm{d}\tilde t\, .
\end{align}
Now, interchanging the order of integration and accordingly adjusting the limits of the double-integral \cite{toda1991statistical},
{\par\nobreak\noindent} 
\begin{align}
	...
	&=
	\int_{0}^{\tf}    \int_{0}^{t''} \Gamma(t''-\tilde{t}) \dot{x} (\tf-t'') \dot x(\tf -\tilde t)  \mathrm{d}\tilde t \mathrm{d}t'' \,,
\end{align}
and subsequentially renaming $t'' \to t$ and $\tilde t \to t'$, we find
{\par\nobreak\noindent} 
\begin{align}
	...
	&=
	\int_{0}^{\tf}    \int_{0}^{t} \Gamma(t-t') \dot{x} (\tf-t) \dot x(\tf -t')  \mathrm{d}t' \mathrm{d}t 
	\,.
\end{align}
Comparing this with the original integral in Eq.~\eqref{G-functional-M}, we notice that 
each path $x$ and the paths defined by
$\hat{x}(t) \equiv \pm x(\tf-t)+ \mathcal{C}_s,$
with some arbitrary constant $ \mathcal{C}_s$, give the same value of the integral term. However, among the possible paths $\hat{x}(t)$, only the one with $\mathcal{C}_s = x(\tf)$ and negative sign of the $x(\tf-t)$ term, is compatible with the equilibrium initial condition and satisfies the constraint $\hat{x}(\tf)=x(\tf)$, such that the remaining contributions to $\la W \ra [x]$ in Eq.~\eqref{G-functional-M}, in particular, the increase of potential and kinetic energy, is also identical for $\hat{x}$ and $x$. 
We conclude that any given trajectory ${x}(t)$ and its time-reversed image $\hat{x}(t)=-{x}(\tf-t)+x(\tf)$ yield the same work. Since the work can be expressed as a quadratic functional of $x$ alone [see Eq.~\eqref{G-functional-M}], we expect only a unique optimum, such that optimal trajectories must satisfy $x^*\equiv\hat{x}$ and, consequentially, must obey time-reversal symmetry.

\subsection{Proof of: Optimality $\Rightarrow$ Symmetry of $\lam$}
Here, we outline the proof of that optimal protocols $\lambda^*$ are time-symmetric.
First, we show how one can generally express $x$ as functional of $\lam$. 
To this end, we take the noise average of the GLE~\eqref{GLE}, which, together with the equilibrium initial condition, yields \eqref{eq:noise-averagedGLE}. To formally solve Eq.~\eqref{eq:noise-averagedGLE} for $x$, we apply a Laplace transformation, $\hat{f}(s) = \int_{0}^{\infty}f(t) e^{-st} \dt$, and make use of the convolution theorem, obtaining
{\par\nobreak\noindent} 
\begin{align}
%
& -\kappa \hat{x}(s) + \kappa \hat\lambda(s)  = m \hat{\ddot x}(s)+	\hat\Gamma (s) \hat{\dot{x}}(s)\nonumber \\
 & = m[s^2\hat{x}(s)-sx(0) - \dot{x}(0)]+\hat\Gamma (s) [s\hat{x}(s) - x(0)] \,.
\end{align}
Recalling $x(0)=0$, $\dot{x}(0)=0$, and solving for $\hat{x}(s)$, we find
{\par\nobreak\noindent} 
\begin{align}
\hat{x}(s)  & = \hat\Phi(s) \, s \hat\lambda(s)\,,
\end{align}
with the response function in Laplace domain 
{\par\nobreak\noindent} 
\begin{align}
  \hat\Phi(s):=\frac{\kappa}{ m s^3+\hat\Gamma (s)s^2  + \kappa s}.   
\end{align}
Noting that $\lam(0)=0$ and thus $s \hat\lambda(s)=[ s \hat\lambda(s)-\lam(0)]$, we transform back to the time domain, and find 
{\par\nobreak\noindent} 
\begin{align}
 x(t) & = \int_{0}^{\infty} \!\!\Phi(t-t') {\dot\lam}(t')\mathrm{d}t'\, .
\end{align}
Assuming causality of the stochastic process implies that any response function must satisfy $\Phi(t<0)=0$, so that we can tighten the integration limits and finally find
{\par\nobreak\noindent} 
\begin{align}
 x(t) & =\int_{0}^{t} \Phi(t-t') {\dot\lam}(t')\mathrm{d}t' \, . 
\end{align}

Now, this expression can be used to replace $x$ in the noise-average of Eq.~\eqref{def:W}, 
directly leading to the following expression of the work as a functional of $\lam$ only, 
{\par\nobreak\noindent} 
\begin{align}\label{eq:W-lamdot-sq}
	\la	W  \ra[\lam]
	&= \frac{\kappa \lf^2}{2} +  \kappa\int_{0}^{\tf} \!  \int_{0}^{t} \! \Phi(t-t') \dot\lam(t)\dot\lam(t') \,\mathrm{d}t'\dt .
\end{align}

Starting from this functional, which is again quadratic in $\lam$, and repeating the analogous steps as we have used before to prove time-reversal symmetry of $x^*$, readily implies that $\lam^*$ possesses time-reversal symmetry, too.

Consistent with our findings, it was shown in Ref.~\cite{naze2022optimal} using linear response theory that optimal protocols of generic Hamiltonian systems in the 
fast-but-weak driving regime are time-symmetric.

\subsection{Proof of: Symmetry of $x$ and $\lam$ $\Rightarrow$ Optimality}
\label{App:Symmetry-x-implies-opt}
Finally, we show that the combined symmetry of $x$ and $\lam$ implies that the process is optimal. The main step of this proof is to derive a generic condition on $x$ for optimality via variational calculus (leading to 
%
Eq.~\eqref{condition_optimalTrajectories}
given in the main text).

To this end, we start from the work functional given in Eq.~\eqref{G-functional}. The variation of $\la W \ra$[x] with respect to variations of $x$ that satisfy $\delta x(0) =\delta x(\tf)=0 $ is generally given by
{\par\nobreak\noindent} 
\begin{align}\label{variation-of-work}
	\delta \la W \ra[x(t),&\delta x(t)] =
	 \int_{0}^{\tf}    \int_{0}^{t} \Gamma(t-t') \dot{x} (t') \delta \dot x(t)  \mathrm{d}t'  \dt
  \nonumber \\
&	+
	 \int_{0}^{\tf}    \int_{0}^{t} \Gamma(t-t')\delta \dot{x} (t') \dot x(t)  \mathrm{d}t'  \dt 	\, .	
\end{align}
The second term of this last expression can be rewritten, by interchanging the integrals, 
and renaming $t\to t'$, $t' \to t$, leading to
{\par\nobreak\noindent} 
\begin{align}
	...	&=	
\int_{0}^{\tf}   \int_{t}^{\tf}  \dot x(t')\Gamma(t'-t)\delta \dot{x} (t)   \mathrm{d}t'  \dt \, .
\end{align}
Now, by using twice the identity: $a=|a|$, $\forall a>0$, we can combine both integral terms of $\delta \la W \ra$ from Eq.~\eqref{variation-of-work}, and find 
{\par\nobreak\noindent} 
\begin{align}
	\delta \la W \ra
	&=
	\int_{0}^{\tf} \! \delta \dot x(t)    \int_{0}^{\tf} \Gamma(|t-t'|) \dot{x} (t') \mathrm{d}t'  \dt \, .
\end{align}
Finally, integration by parts
 and using that, per definition, $\delta x(0) =\delta x(\tf)=0 $,
we obtain
{\par\nobreak\noindent} 
\begin{align}
	\delta \la W \ra
	&=
\int_{0}^{\tf} \! \delta  x(t)  \frac{\mathrm{d}}{\mathrm{d}t} \left[\int_{0}^{\tf} \Gamma(|t-t'|) \dot{x} (t') \mathrm{d}t'  \right] \dt \, .
\end{align}
Thus, trajectories that satisfy the condition
{\par\nobreak\noindent} 
\begin{align}\label{condition_optimalTrajectories-M}
	 \frac{\mathrm{d}}{\mathrm{d}t} \left[\int_{0}^{\tf} \Gamma(|t-t'|) \dot{x}(t') \mathrm{d}t'  \right] = 0,~~~\forall t\in [0,\tf]
\end{align}
automatically fulfill $\delta \la W \ra[x(t),\delta x(t)]=0$ for all variations with $\delta x(0) =\delta x(\tf)=0 $, which is a characteristic property of optimal solutions. [The condition Eq.~\eqref{condition_optimalTrajectories-M} is identical to Eq.~\eqref{condition_optimalTrajectories} given in the main text.]

For sake of completeness, we note that, starting from Eq. \eqref{eq:W-lamdot-sq}, analogous steps yield the analog condition 
{\par\nobreak\noindent} 
\begin{align}\label{condition_optimalProtocols}
	 \frac{\mathrm{d}}{\mathrm{d}t} \left[\int_{0}^{\tf} \Phi(|t-t'|) \dot{\lam} (t') \mathrm{d}t'  \right] = 0.~~~\forall t\in [0,\tf] \, ,
\end{align}
for optimal protocols.

The second step of this part of the proof is to show that symmetric control processes automatically satisfy the condition \eqref{condition_optimalTrajectories-M}. 
We start with assuming that a given protocol is time-symmetric, i.e.,
{\par\nobreak\noindent} 
\begin{align}
\lf=\lam(t) +\lam(\tf-t).\end{align} Substituting in this equation $\lam(t)$ and $\lam(\tf-t)$ using the noise-averaged GLE \eqref{eq:noise-averagedGLE}, yields
{\par\nobreak\noindent} 
\begin{align}\label{eq:intermediate-step-proof}
& \lf =	 \frac{1}{\kappa} \int_0^{t}\Gamma(t-t') \dot{x}(t')\mathrm{d}t'
+ \frac{m}{\kappa} \ddot{x}(t)
+ x(t) 
\nonumber \\
&+\frac{1}{\kappa}  \int_0^{\tf-t}\Gamma(\tf-t-t') \dot{x}(t')\mathrm{d}t'
+ \frac{m}{\kappa} \ddot{x}(\tf-t)
+x(\tf-t)
 \,.
\end{align}
Now, 
transforming the coordinates of the second integral ($t'=\tf-t''$)
and assuming that the trajectory $\x$ also obeys time-reversal symmetry, which implies ${x}(t)+x(\tf-t)=x(\tf)$ and $\ddot{x}(t)+\ddot{x}(\tf-t)=0$, we find
{\par\nobreak\noindent} 
\begin{align}
 \kappa[\lf - x(\tf)]=& 
 \int_0^{t}\Gamma(t-t') \dot{x}(t')\mathrm{d}t'
 \nonumber \\&
 +\int_{t}^{\tf}\Gamma(t''-t) \dot{x}(\tf-t'')\mathrm{d}t'' 
	 \,.
 \end{align}
Finally, using $\dot{x}(\tf-t)=\dot{x}(t)$ which is implied by the time-reversal symmetry of $x$, and again the identity: $a=|a|$, $\forall a>0$, 
we find 
{\par\nobreak\noindent} 
\begin{align}\label{eq:intermediate-step-proof-2}
 \kappa[\lf - x(\tf)]=& 	\int_0^{t}\Gamma(|t-t'|) \dot{x}(t')\mathrm{d}t'\nonumber \\
	&+\int_{t}^{\tf}\Gamma(|t-t''|) \dot{x}(t'')\mathrm{d}t'' 
	 \,
 \end{align}
which readily implies 
{\par\nobreak\noindent} 
\begin{align}\label{eq:intermediate-step-proof-3-M}
	\int_0^{\tf}\Gamma(|t-t'|) \dot{x}(t')\mathrm{d}t'
	& = \kappa[\lf - x(\tf)] \, ,
\end{align}
as given in Eq.~\eqref{eq:intermediate-step-proof-3} in the main text.
It is easy to see that this equality, which is as we have just shown fulfilled by all processes with time-symmetric $x$ and $\lam$, readily implies that condition \eqref{condition_optimalTrajectories-M} is fulfilled.  
This further implies that time-symmetric processes are generally optimal. We note that this proof also holds in the overdamped limit.

We note that the work required to translate the trap can be reduced if additional real-time information about the actual particle position is acquired, e.g., by performing a measurement at $t=0$ and modifying the protocol at $t>0$ according to the measurement outcome~\cite{abreu2011extracting}. In the presence of such measurements, the symmetry property discussed here is lost.

\section{Model and theoretical prediction of optimal protocol in viscoelastic fluid}
\label{sec:protVE}
Here, we derive the optimal protocols for the particle in a Maxwell fluid described by the overdamped limit of the 
GLE~\eqref{GLE} with $\Gamma(t-t') = 2\gamma \delta (t-t') + \kappa_b e^{-(t-t')/\tau_b},$ 
and $\langle \nu(t)\nu(t')\rangle = 2k_\mathrm{B}T [\gamma \delta(\Delta t)+\gamma_b e^{-|\Delta t|/\tau_b}]$, with the bath stress-relaxation time $\tau_b$. 
Equivalently, the dynamics of $X$ can be described by Eqs.~\eqref{LEx} and \eqref{LEy}, as given in the main text.\\
As a first step, we express the work as a functional of $\x$ and $x_b := \langle X_b\rangle$, only. To this end, 
we use the noise average of Eqs.~\eqref{LEx} and \eqref{LEy}, which is
{\par\nobreak\noindent} 
\begin{align}
\tau_p\,\dotx &= -(k+1)\,\x + \xb + k\,\lambda , \label{eq:x}   \\
\tau_b\,\dotxb &= -\xb +\x \label{eq:xb} \,.
\end{align}
Solving Eq.~\eqref{eq:x} for $\lam$ and substituting the result, we can rewrite the expression for the work $W$ as
{\par\nobreak\noindent} 
\begin{align}
\la W \ra[\x,\xb]  
  =& \frac{\kappa}{k^2}\int_{0}^{\tf}\! \mathrm{d}t 
 \left[
 \tau_p^2\ddotx\dotx 
+ \tau_p \ddotx \x - \tau_p \ddotx \xb - \tau_p \dotx \dotxb
\right.
\nonumber \\
 & \left.
   + \tau_p (k+1)\dotx^2
   +(k+1)\dotx \x 
  -(k+1)\dotx \xb
\right.
\nonumber \\
 & \left.
- \x\dotxb  
+ \dotxb \xb \right].
\end{align}
This can be simplified by explicitly integrating all terms of the form $\int_{0}^{\tf} \, \dotx \x  \mathrm{d}t= \frac{1}{2}[\x^2]_{0}^{\tf}$, and rewriting by partial integration some of the other terms like $\int_{0}^{\tf}  \,\dotx \xb \mathrm{d}t = \frac{1}{2}[\x \xb]_{0}^{\tf}-\int_{0}^{\tf}  \x \dotxb \mathrm{d}t $; leading to 
{\par\nobreak\noindent} 
\begin{align}\label{Wuv}
	\frac{k}{\kappa}\, \la W \ra[x,\xb]
 =& \int_{0}^{\tf} \mathrm{d}t 
 (\tau_p \dotx^2 -\dotx {\xb})
  \nonumber
 \\
 &+ \frac{1}{2k}\left[
 {\tau_p^2}\dotx ^2 
 +\left(k+1\right)\x^2
 +\xb^2\right]_{0}^{\tf}
  \nonumber
 \\
 &+ \frac{1}{k}\left[\tau_p\,\dotx \x -\tau_p\dotx \xb- \x {\xb}\right]_{0}^{\tf}\,.
\end{align}
\\
Next, to minimize the work functional~\eqref{Wuv}, we incorporate as a constraint the noise-averaged Eq.~\eqref{LEy} via a Lagrange multiplier $\Lambda(t)$, giving rise to the cost functional (Lagrangian)
{\par\nobreak\noindent} 
\begin{align}\label{eq:costfunctional}
    \mathcal{L}[\x,\dot \x,\dot \xb] =  [\tau_p\,\dotx^2- \dotx {\xb}]+ \Lambda(t)[\tau_b\,\dotxb+\xb -\x].
\end{align} 
%
%
For this cost functional, 
the Euler-Lagrange equations
{\par\nobreak\noindent} 
\begin{align}
	\frac{\partial \mathcal{L}}{\partial \x} - \frac{\mathrm d}{\mathrm dt}\frac{\partial \mathcal{L}}{\partial \dotx} &= 0 ,\\
		\frac{\partial \mathcal{L}}{\partial \xb} -\frac{\mathrm d}{\mathrm dt}\frac{\partial \mathcal{L}}{\partial \dotxb} &= 0 ,\\
\frac{\partial \mathcal{L}}{\partial \Lambda} -\frac{\mathrm d}{\mathrm dt}\frac{\partial \mathcal{L}}{\partial \Lambda} & = 0 ,
\end{align}
yield the set of three linear, coupled second-order differential equations: $2 \tau_p \ddotx = \dotxb -\Lambda $ and $\tau_b\dot \Lambda =-\dotx +\Lambda$, 
as well as Eq.~\eqref{eq:xb}, which we included as dynamical constraint. Introducing the variable $v := \dotx,$ 
we 
can rewrite the latter three equations as 
the set of four coupled linear, first-order differential equations
{\par\nobreak\noindent} 
\begin{align}
\dot{\boldsymbol{z}}=&\mathbb{A} \boldsymbol{z},~~~~\text{with}~~~~
\boldsymbol{z}=    \begin{pmatrix}
    \x &
    v &
    \xb &
    \Lambda
    \end{pmatrix}^T,\\
\mathbb{A}=  &  \begin{pmatrix}
0 & 1 & 0 & 0 \\
\frac{1}{2 \tau_b \tau_p} & 0 & -\frac{1}{2 \tau_b \tau_p} & -\frac{1}{2 \tau_p} \\
\frac{1}{\tau_b} & 0 & -\frac{1}{\tau_b} & 0 \\
0 & -\frac{1}{\tau_b} & 0 & \frac{1}{\tau_b}
\end{pmatrix} .
\end{align}	 
The solution of this equation is $\boldsymbol{z}=\boldsymbol{z}(0)e^{\mathbb{A}t}$.
In combination with the initial conditions $\x(0) = 0,\xb(0)=0$, this yields the optimal 
solution
\begin{widetext}
{\par\nobreak\noindent} 
\begin{align}\label{eq:optimal-solutionx}
\x^*(t) &= \frac{1}{2(\tau_b +\tau_p)}
\left[
2 \tau_p \mathcal{C}_1 t + \tau_b(\tau_b + t)\mathcal{C}_2 + \tau_b^2 \left\{ -\mathcal{C}_2 \cosh{\frac{\sqrt{\tau_b+\tau_p}}{\tau_b\sqrt{\tau_p}}t} + \frac{\sqrt{\tau_p}(2\mathcal{C}_1-\mathcal{C}_2)}{\sqrt{\tau_b+\tau_p}}\sinh{\frac{\sqrt{\tau_b+\tau_p}}{\tau_b\sqrt{\tau_p}}t}\right\} 
\right] 
\\
\xb^*(t) &= \frac{1}{2(\tau_b +\tau_p)}
\left[
2 \tau_p \mathcal{C}_1  +
\tau_b \mathcal{C}_2
+
\tau_b(2\mathcal{C}_1-\mathcal{C}_2) \cosh{\frac{\sqrt{\tau_b+\tau_p}}{\tau_b\sqrt{\tau_p}}t} 
-
\tau_b \frac{\sqrt{\tau_b+\tau_p}}{\sqrt{\tau_p}} \sinh{\frac{\sqrt{\tau_b+\tau_p}}{\tau_b\sqrt{\tau_p}}t}
\right] 
\\
\lambda^*(t) &= 
\frac{1}{2(\tau_b +\tau_p)k}
\left[
2 \tau_p (\tau_b + \tau_p+k t)\mathcal{C}_1 + \tau_b(\tau_b + \tau_p+\tau_b k + k t)\mathcal{C}_2 
\right.
\nonumber \\
&
~~~
\left.
- \tau_b (\tau_b+\tau_p +\tau_b k) \mathcal{C}_2 \cosh{\frac{\sqrt{\tau_b+\tau_p}}{\tau_b\sqrt{\tau_p}}t} + \frac{\tau_b\sqrt{\tau_p}(\tau_b + \tau_p+\tau_b k)(2\mathcal{C}_1-\mathcal{C}_2)}{\sqrt{\tau_b+\tau_p}}\sinh{\frac{\sqrt{\tau_b+\tau_p}}{\tau_b\sqrt{\tau_p}}t} 
\right] \, . \label{eq:lambdaoptfull}
\end{align}
\end{widetext}
Note that in the last step we have used Eq.~\eqref{eq:x} to obtain 
$\lambda^*$ from 
$\x^*$ and $\xb^*$. 
Thus, although the equation of motion of the particle is linear in $X$, as it is for the Markovian case, the optimization now leads to fully nonlinear solutions. In particular, there is no ``steady-state-like'' regime anymore (where particle and trap move at a constant distance with constant speed).

The solutions obtained by this optimization procedure given in Eqs.~\eqref{eq:optimal-solutionx}-\eqref{eq:lambdaoptfull}
still depend on two unknown parameters $\mathcal{C}_1$ and $\mathcal{C}_2$. 
We remark that the long-time limit of Eqs.~\eqref{eq:optimal-solutionx}-\eqref{eq:lambdaoptfull} reveals that these unknowns encode the initial jump of $\dotx$ and the initial value of the Lagrange-multiplier:
$\mathcal{C}_1=\dotx(0^+),~~\mathcal{C}_2=\Lambda(0^+)$, which depend on all parameters $\{\lf,\tf,\tau_b,\tau_p,\kappa_b,k\}$.

Usually, these unknowns must be determined by a secondary minimization. Concretely, by inserting $\x^*$ and $\xb^*$, the work \eqref{Wuv} can be expressed as function of $\mathcal{C}_1,\mathcal{C}_2$. Then, minimizing $\la W\ra (\mathcal{C}_1,\mathcal{C}_2)$ with respect to $\mathcal{C}_1$ yields the optimal value of $\mathcal{C}_1(\mathcal{C}_2)$ as a function of $\mathcal{C}_2$. Inserting this result, subsequently minimizing $\la W \ra(\mathcal{C}_2)$ with respect to $\mathcal{C}_2$, yields the optimal $\mathcal{C}_2$. %
For the Maxwell fluid, this involves very cumbersome and nested expressions. %
However, this secondary minimization can be entirely avoided by making use of the symmetry property of the optimal mean trajectory and protocol,
which we have proven on general grounds. Postulating time-reversal symmetry of $x^*$ and $\lambda^*$, immediately leads to analytical expressions for $\mathcal{C}_{1,2}(\lf,\tf,\tau_b,\tau_p,\kappa_b,k)$; resulting in closed-form solutions $\{x^*,\xb^*,\lam^*\}$.

As a consistency check, we have verified that the solutions obtained via the secondary minimization (without postulating the symmetry) match the analytical expressions found via the symmetry.

\section{Machine learning algorithm}
\label{sec:ML}
We chose an architecture and training procedure similar to Ref.~\cite{whitelam2023demon}. We used feed-forward fully connected Deep Neural Networks (DNN) that take a single scalar input, and return a single scalar output value, and have three hidden layers of 4,4,10 nodes, as illustrated in Fig.~\ref{fig:ML}.
The set $\theta$ of weights and biases of all nodes parameterizes the DNN. We apply a ReLU activation function to the input layer, and tanh activation function to all other layers. 
The series of time steps $t_i \in \{\Delta t,2\Delta t,...,\tf-\Delta t\}$ with $\Delta t>0$ and $0<t_i<\tf$, is sequentially given  as input to the DNN, which returns as sequential output the protocol values $\lambda(t_i)$.
The generated protocol is complemented by the values $\lambda(0)=\lam_0$ and $\lambda(\tf)=\lf$ which are fixed by the boundary conditions.
Thus, there is a direct mapping between $\theta$ and a protocol $\lambda$.
The DNN is initialized by setting all its parameters $\theta$ to zero, so that it generates protocols that are zero for all time steps $t_i<\tf$ and abruptly jump to $\lf$ at $\tf$. The work corresponding to such protocol is $\langle W \rangle_{0} =  \lf^2/2\,(\gamma_0 / \tau_0)$, and the asymmetry is $\langle A_x + A_\lam \rangle_{0}  = \lf^2/2 \,(\Delta t/\tf)$.
The training by a Monte-Carlo algorithm is done as follows. At each training step $n>0$, 
a copy of the DNN $\theta_{n-1}$ is generated, and all parameters of the copy are perturbed by a Gaussian noise with zero mean and variance $\sigma = 0.03$, leading to $\theta'_n$. For the protocol generated by $\theta'_n$, we then numerically compute (by solving the Langevin equation) the new tentative value of the objective $\phi_{n}$, for which we choose either the average work ($\phi_n \equiv \langle W \rangle_n$) or asymmetry parameter ($\phi_n \equiv \langle A_x + A_\lam\rangle_n$). 
If $\phi'_n < \phi_{n-1}$
the DNN is replaced by the perturbed copy ($\theta_n\leftarrow \theta'_n$), and $\phi_n=\phi'_n$. Otherwise the copy DNN is rejected, and $\phi_n=\phi_{n-1}$.
Expressing times in units of $\tau_0$, and space in units of $\lf$, de-dimensionalises the dynamical equations, allowing us to keep $\kappa$ and $\tau_0$ generic. We chose $\tf=\tau_0$, and a temporal discretization of $\Delta t/\tf = 10^{-3}$. Note that the chosen network architecture and parameters provide robust results and fast convergence, but we have not performed a systematic optimization of the hyperparameters (i.e., the number of nodes and layers, choice of activation function and temporal discretization).

\nolinenumbers
\newpage

\nolinenumbers
\bibliography{bib}

\end{document}